\documentclass[12pt]{article}
\usepackage[latin1]{inputenc}
\usepackage{amsfonts,amssymb,amsmath, epsfig}
\usepackage{color,graphicx,graphics}
\usepackage{amsthm}
\usepackage{amsmath,amstext,amssymb,amsfonts, amscd}
\usepackage[lofdepth,lotdepth]{subfig}
\usepackage{xcolor}
\usepackage{hyperref}

\usepackage{stix}

\hypersetup{
colorlinks=true,
linkcolor=blue,
citecolor=blue,
urlcolor=blue,
}

\def\Box{\leavevmode\vbox{\hrule
     \hbox{\vrule\kern4pt\vbox{\kern4pt}%
           \vrule}\hrule}}
\def\blackbox{\leavevmode\vrule height 5pt width 4pt depth 0pt\relax}
\def\endproof{\null\hfill {$\blackbox$}\bigskip}

\def\paragraph#1{{\bf #1\ }}

\newtheorem{lemma}{Lemma}[section]

\newtheorem{proposition}[lemma]{Proposition}

\newtheorem{remark}{Remark}[section]

\title{Pedestrian models with congestion effects} 
\author{Pedro Aceves-S\'anchez$^{(1)}$, Rafael Bailo$^{(2)}$, Pierre Degond$^{(3)}$, Zo\'e Mercier$^{(4)}$} 
\date{} 
\begin{document}

\maketitle

\vspace{0.5 cm}

\begin{center}
$^{(1)}$ Department of Mathematics, University of Arizona, USA \\
{\sffamily pedroas@math.arizona.edu}
\end{center}

\begin{center}
$^{(2)}$ Mathematical Institute, University of Oxford, United Kingdom \\
{\sffamily bailo@maths.ox.ac.uk}
\end{center}

\begin{center}
$^{(3)}$ Institut de Math\'ematiques de Toulouse ; UMR5219 \\
Universit\'e de Toulouse ; CNRS \\
UPS, F-31062 Toulouse Cedex 9, France\\
{\sffamily pierre.degond@math.univ-toulouse.fr}
\end{center}

\begin{center}
$^{(4)}$ Institut de M\'ecanique des Fluides de Toulouse; UMR CNRS/INP-UPS 5502; \\
F-31400 Toulouse Cedex, France \\
{\sffamily zoe.mercier@imft.fr}
\end{center}

\vspace{0.5 cm}
\begin{abstract}
We study the validity of the dissipative Aw-Rascle system as a macroscopic model for pedestrian dynamics. The model uses a congestion term (a singular diffusion term) to enforce capacity constraints in the crowd density while inducing a steering behaviour. Furthermore, we introduce a semi-implicit, structure-preserving, and asymptotic-preserving numerical scheme which can handle the numerical solution of the model efficiently. We perform the first numerical simulations of the dissipative Aw-Rascle system in one and two dimensions. We demonstrate the efficiency of the scheme in solving an array of numerical experiments, and we validate the model, ultimately showing that it correctly captures the fundamental diagram of pedestrian flow.
\end{abstract}

\medskip
\noindent
{\bf Acknowledgements:} RB was supported by the Advanced Grant Nonlocal-CPD (Nonlocal PDEs for Complex Particle Dynamics: Phase Transitions, Patterns and Synchronization) of the European Research Council Executive Agency (ERC) under the European Union's Horizon 2020 research and innovation programme (grant agreement No.~883363) and by the EPSRC grant EP/T022132/1 ``Spectral element methods for fractional differential equations, with applications in applied analysis and medical imaging". PD holds a visiting professor association with the Department of
Mathematics, Imperial College London, UK.


\vskip 0.4cm

\setcounter{equation}{0}
\section{Introduction}

There exist many mathematical models for the motion of people in crowds; however, these models often struggle to enforce \textit{capacity constraints}: the fact the density of people in a confined space cannot physically exceed a certain threshold. In this work, we explore the so-called \textit{dissipative Aw-Rascle} system \cite{CFZ2022,CGZ2023,CNP2022} as a macroscopic description of pedestrian motion that enforces a capacity constraint and introduces a steering behaviour. For the first time, we explore this system from a modelling perspective, and perform the first numerical simulations in one and two spatial dimensions. To that end, we introduce a structure-and-asymptotic-preserving numerical scheme that can simulate the model efficiently across regimes.

Pedestrian dynamics has for some time been a topic of practical relevance. The behaviour of humans moving in crowds is complex and has been studied at length, from early engineering studies\cite{Fruin1971,HW1958,Older1968} to modern policymaking  \cite{HassKlau2015,HCM2016}, with perspectives of both design \cite{HBJ2005} and safety \cite{HFV2000}. Experimental work in crowd motion has been performed to understand emergent features, including the \textit{fundamental diagram} \cite{SSK2005,Weidmann1993}, lane formation \cite{HMF2001,MHG2009}, or stop-and-go waves \cite{HJA2007}. Some studies have tackled circumstantial effects such as fatigue \cite{LFZ2016} or cluster formation \cite{MGM2012}. Others reveal the different features observed as pedestrians encounter intersections \cite{HBJ2005}, bottlenecks \cite{KGS2006}, or counterflows \cite{KGK2006}.

The study of pedestrians has also drawn attention in the mathematical community. \textit{Agent-based models} have been applied to crowds; a review can be found in \cite{BD2011}. Some of these models are based on alignment and force principles, \cite{HM1995} which are also used in broader particle dynamics modelling \cite{DCB2006,Reynolds1987,Reynolds1999} because they lead to emergent properties such as phase transitions \cite{BCC2016,VCB1995}. More recently, \textit{rational pedestrian models} have been introduced \cite{BCD2018,DAM2013,MHT2011,STK2023}, which incorporate the roles of visual stimuli and anticipation \cite{CVB1995,SKS2000} into microscopic dynamics.

The macroscopic perspective is an alternative route to crowd modelling; \cite{BB2011,BBC2011} are recent reviews. Early hydrodynamic approaches \cite{Henderson1971,Henderson1974} followed on the footsteps of vehicular traffic modelling \cite{LW1955b,Richards1956}, in a framework where only the features of the crowd as a whole (such as the pedestrian density or the flow through a corridor), and not those of the individual, are of interest. Some macroscopic models are derived from microscopic principles trough the kinetic point of view \cite{BB2010,DAM2013,DAP2013,Helbing1992}, and the study of the correspondence between microscopic dynamics and the macroscopic scale is a very active field in general \cite{BAP2011,CKP2014,CPT2011}, though a fully rigorous analysis can only be performed in some cases \cite{BCC2012,CDW2013}. Sometimes, corresponding microscopic models are found after a macroscopic model has been proposed \cite{AKR2002,BDM2014,DT2020}. Models can also be prescribed directly \cite{CMW2016,CR2005,Hughes2002,Hughes2003}, an approach which has been successful in encoding physical and social principles. Other approaches to modelling include cellular automata \cite{BKS2001}, mean-field games \cite{Dogbe2010,LW2011}, and multi-scale models \cite{BPT2012,CPT2011}.

This work will also focus on a macroscopic model inspired by vehicular traffic: we study the validity of the dissipative Aw-Rascle system as a model for pedestrian dynamics. The system was studied in \cite{CFZ2022,CGZ2023,CNP2022}, based on the model of \cite{AR2000,BDD2008}. The model incorporates a singular \textit{congestion} term. The congestion effect is negligible everywhere except in the regions where the crowd density approaches an upper bound (the capacity), and it ensures that said bound is not exceeded; this locality is achieved by prescribing an asymptotic strength to the congestion. In order to handle the stiffness of the problem efficiently, while guaranteeing that the capacity bound is satisfied by the numerical solution, we adapt a scheme which has been successfully applied to the compressible Euler equations \cite{DHN2011,DT2011}, which is implicit and \textit{asymptotic preserving} \cite{HJL2017,Jin2012}.

The work is organised as follows: Section \ref{sec:model} introduces the pedestrian model; Section \ref{sec:scheme} constructs an implicit, structure-and-asymptotic-preserving, finite-volume scheme; Section \ref{sec:experiments} presents numerical experiments (also available online \cite{ABDPrepWeb} in interactive form, and in a permanent repository \cite{ABDPrepFig} as videos) that validate the scheme and showcase its properties; finally, conclusions are presented in Section \ref{sec:outlook}. Some additional figures and details about the two-dimensional implementation of the numerical schemes are presented in the appendices.

\setcounter{equation}{0}
\section{The pedestrian model with congestion effects}
\label{sec:model}

This work considers \textit{hydrodynamic models} where we model the macroscopic \textit{density of pedestrians} within a crowd, $\rho(t,\mathbf{x})\in{\mathbb R}$, as well as the \textit{physical velocity} of the crowd, $\mathbf{u}(t,\mathbf{x})\in{\mathbb R}^2$. An elementary assumption is \textit{conservation of mass}: pedestrians may not enter or exit the domain except on its boundary. Therefore, the evolution in time of the density must obey the \textit{continuity equation}
\begin{align*}
	\partial_t\rho + \nabla \cdot \mathbf{J} = 0,
	\quad \mathbf{J} =: \rho \mathbf{u}, 
	\quad t>0, \quad \mathbf{x}\in\Omega\subseteq{\mathbb R}^2.
\end{align*}

\subsection{Derivation of the model}

The modelling task is reduced to the design of an evolution law for the velocity $\mathbf{u}$. To do so, we draw ideas from the traffic modelling literature, beginning with the one-dimensional model of Aw and Rascle \cite{AR2000} in its \textit{rescaled} and \textit{modified} form \cite{BDD2008}:
\begin{align}\label{eq:RMAR}
	\begin{cases}
		\partial_t \rho + \partial_x ( \rho u ) = 0,                          \\
		( \partial_t + u\partial_x ) ( u + \varepsilon p(\rho) ) = 0, \\
		p(\rho) = \big( \rho^{-1} - \rho_{\text{max}}^{-1} \big)^{-\gamma},
	\end{cases}
	\quad t>0, \quad x \in \Omega \subseteq {\mathbb R},
\end{align}
for some exponent $\gamma>0$ and some \textit{asymptotic parameter} $\varepsilon>0$. The unknowns $\rho$ and $u$ are now the density and velocity of traffic, and $\rho_{\text{max}}>0$ is the \textit{capacity}: the ``bumper-to-bumper'' density, where there is no longer any free space on the road. The function $p(\rho)$ is known as the \textit{pseudo-pressure} (and other non-decreasing functions of the density could be considered instead, so long as $p(0)=0$). The original Aw-Rascle model \cite{AR2000} was proposed in order to address the historical shortcomings of second-order models \cite{Daganzo1995,Payne1971,Whitham1974}, where information could propagate forward faster than traffic itself, a critical modelling flaw. Contrary to this, the characteristic speeds of \eqref{eq:RMAR} are given by $u - \rho \varepsilon p'(\rho)$ and $u$, ensuring that information only propagates backwards relative to the drivers.

The model is better understood by defining the \textit{desired velocity} $w =: u + p(\rho)$, which is advected by \eqref{eq:RMAR}, and thus constant along characteristics (i.e., intrinsic to each driver). Rewriting $u = w - p(\rho)$, we see that the physical speed will equal the desired speed when the traffic density is zero, and then decrease as the density increases. The pseudo-pressure $p(\rho)$ thus models a basic but fundamental principle of traffic: drivers are free to travel at their preferred speed only if the density of traffic is sufficiently low. As the density increases, the typical separation between vehicles decreases, and drivers have to account for their \textit{reaction time} (as well as everybody else's) by reducing their speed.

\begin{figure}[h!]
	\centering
	\includegraphics{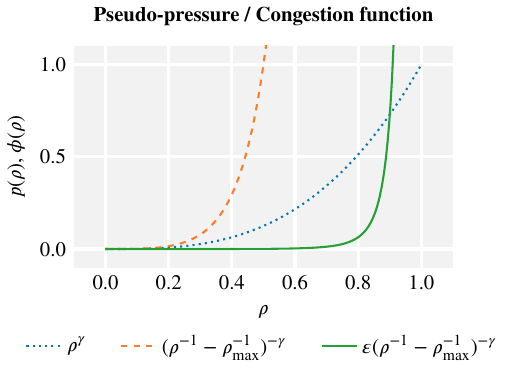}
	\caption{
	Typical forms of the pseudo-pressure $p(\rho)$ and congestion function $\phi(\rho)$.
	$\gamma=3$, $\varepsilon=10^{-3}$.
	}
	\label{fig:pseudo_pressure}
\end{figure}

The original formulation \cite{AR2000} employs $p(\rho)=\rho^\gamma$ for $\gamma>0$, a choice that facilitates the analysis of the model. The modified form in \eqref{eq:RMAR} was introduced in \cite{BDD2008}. This pseudo-pressure behaves like $\rho^\gamma$ in the range $\rho \ll \rho_{\text{max}}$ (see Fig. \ref{fig:pseudo_pressure}), but the singularity at $\rho_{\text{max}}$ enforces a \textit{capacity constraint}: solutions to \eqref{eq:RMAR} satisfy $\rho(t,x) \leq \rho_{\text{max}}$. The work also introduces an asymptotic rescaling $p \rightarrow \varepsilon p$, which renders congestion effects negligible outside of congested regions. This introduces a switching behaviour: drivers only reduce their speed if the local density is close to capacity. This behaviour renders the dynamics more realistic, and is the main feature that we seek to incorporate into our pedestrian model.

In order to model pedestrians in general, \eqref{eq:RMAR} must be made two-dimensional. However, the pseudo-pressure term is a scalar, and needs to be suitably adapted. A natural idea is to replace it by the gradient of a \textit{congestion function}, $ \nabla  \phi(\rho)$; this makes the system consistent in higher dimensions. We thus arrive at the \textit{pedestrian model with congestion effects}:
\begin{align}\label{eq:H}
	\begin{cases}
		\partial_t \rho + \nabla \cdot ( \rho \mathbf{u} ) = 0,                                         \\
		( \partial_t + \mathbf{u}\cdot \nabla  ) ( \mathbf{u} + \varepsilon  \nabla  \phi (\rho) ) = 0, \\
		\phi(\rho) = \big( \rho^{-1} - \rho_{\text{max}}^{-1} \big)^{-\gamma},
	\end{cases}
	\quad t>0, \quad \mathbf{x} \in \Omega \subseteq {\mathbb R}^2,
\end{align}
in its \textit{hydrodynamic} form. This system is sometimes referred to as the \textit{dissipative Aw-Rascle system} \cite{CFZ2022,CGZ2023,CNP2022}. Our congestion function takes the same form as the pseudo-pressure from \cite{BDD2008}, but any non-decreasing function of the density may be used instead, so long as it maintains the singularity at $\rho_{\text{max}}$. In practice, the congestion function must be fitted from data, but we will leave this question aside in the present work. 

Once again the model is better understood in terms of the \textit{desired velocity}, the quantity advected by \eqref{eq:H}, now given by $ \mathbf{w}  =: \mathbf{u} + \varepsilon  \nabla \phi(\rho)$. The system can be rewritten (at least for smooth solutions) as
\begin{align}\label{eq:AD}
	\begin{cases}
		\partial_t \rho + \nabla \cdot (\rho  \mathbf{w} ) = \varepsilon \nabla \cdot (\rho  \nabla  \phi(\rho) ), \\
		\partial_t  \mathbf{w}  + ( \mathbf{w}  - \varepsilon  \nabla  \phi (\rho)) \cdot  \nabla   \mathbf{w}  = 0,       \\
		\phi(\rho) = \big( \rho^{-1} - \rho_{\text{max}}^{-1}\big)^{-\gamma},
	\end{cases}
	\quad t>0, \quad \mathbf{x} \in \Omega \subseteq {\mathbb R}^2,
\end{align}
which we call the \textit{advection-diffusion} form of the model. This form reveals the two main features of the model:

The first equation enforces the \textbf{capacity bound}. It is a continuity equation for the density, $\rho$, with respect to the desired velocity, $ \mathbf{w} $, together with a non-linear, density-dependent diffusion term. We observe that the non-decreasing property of $\phi$ is essential to ensure the diffusion is well-posed. Moreover, the singular character of $\phi'$ (see Fig. \ref{fig:pseudo_pressure}) enforces the constraint $\rho < \rho_{\text{max}}$, since the strength of the diffusion increases arbitrarily as the density approaches the bound.

The second equation introduces a \textbf{steering behaviour}, a new feature which was not present in \eqref{eq:RMAR}. It advects $ \mathbf{w} $ with a velocity equal to itself minus a correction term; this term drives the advection along the negative gradient of the density, away from congestion. The advective (physical) velocity, $\mathbf{u} =  \mathbf{w}  - \varepsilon \phi'(\rho)  \nabla \rho$, models not only a reduction of the speed in crowded areas but also the steering: pedestrians who face an incoming crowd will reduce their speed, but also seek to move \textit{around} the congestion. This behaviour is fundamental in such aspects of pedestrian dynamics as \textit{collision avoidance} \cite{Batty1997,Gigerenzer2008} and \textit{lane formation} \cite{HMF2001,MHG2009}; its relevance will be demonstrated in Section \ref{sec:experiments}.

Finally, we remark that, while the present derivation of the model is based on phenomenological considerations, another derivation from a system of interactive active particles is also possible and will be exposed in Section \ref{sec:deriv_particles}. 

Before proceeding, we rewrite \eqref{eq:H} in \textit{conservative form}, more suited to numerical simulation. We multiply the density equation by $ \mathbf{w} $, the velocity equation by $\rho$, and sum both equations. This readily simplifies to
\begin{align}\label{eq:C}
	\begin{cases}
		\partial_t \rho + \nabla \cdot  \mathbf{q}  = \varepsilon \nabla \cdot \big( \rho  \nabla  \phi(\rho) \big),                                   \\
		\partial_t  \mathbf{q}  + \nabla \cdot \Big(\frac{ \mathbf{q}  \otimes  \mathbf{q} }{\rho} \Big)  = \varepsilon \nabla \cdot \big(  \nabla  \phi(\rho) \otimes  \mathbf{q}  \big), \\
		\phi(\rho) = \big( \rho^{-1} - \rho_{\text{max}}^{-1}\big)^{-\gamma},
	\end{cases}
	\quad t>0, \quad \mathbf{x} \in \Omega \subseteq {\mathbb R}^2,
\end{align}
where $ \mathbf{q}  =: \rho \mathbf{w} $ is the \textit{desired momentum} and where, for a tensor $A = (A_{ij})_{i,j=1, 2} = A(x)$, $\nabla \cdot A$ denotes the vector of components $(\nabla \cdot A)_j = \sum_{i=1}^2 \partial_{x_i} A_{ij}$.

\subsection{Limiting behaviour}
\label{sec:limit_explanation}

It is possible to consider the behaviour of the model in the $\varepsilon\rightarrow 0$ limit. If the congestion function $\phi$ were not singular, we would expect the terms $ \nabla \phi(\rho)$ to scale like $\mathcal{O}(\varepsilon)$. In the limit, the second order terms would vanish, and model \eqref{eq:C} would formally become the so-called \textit{pressureless gas dynamics system} \cite{Bouchut1994}:
\begin{align}\label{eq:PGD}
	\begin{cases}
		\partial_t \rho + \nabla \cdot  \mathbf{q}  = 0, \\
		\partial_t  \mathbf{q}  + \nabla \cdot \Big( \frac{ \mathbf{q} \otimes \mathbf{q} }{\rho} \Big) = 0,
	\end{cases}
	\quad t>0, \quad \mathbf{x} \in \Omega \subseteq {\mathbb R}^2.
\end{align}
In this setting, using $ \mathbf{q} =\rho \mathbf{w} $ and assuming sufficient regularity, it can be deduced that the desired velocity satisfies Burgers' equation,
\begin{align*}
	\partial_t  \mathbf{w}  + ( \mathbf{w}  \cdot  \nabla )  \mathbf{w}  = 0,
	\quad t>0, \quad \mathbf{x} \in \Omega \subseteq {\mathbb R}^2,
\end{align*}
which is known to develop shocks in finite time. The shocks in $ \mathbf{w} $ correspond to the appearance of Dirac deltas in the density, and thus we see that the solution to \eqref{eq:PGD} has a tendency to become unbounded.

However, the singular character of the congestion function alters this behaviour completely. The densities which solve \eqref{eq:C} are always bounded above by $\rho_{\text{max}}$. Thus,  this bound is expected to persist in the limit. Since the term $\phi(\rho)$ is finite wherever $\rho$ is below the congestion bound, we will observe $\varepsilon \phi(\rho) \rightarrow 0$ away from the congestion. In congested areas, however, it is expected that $\varepsilon \phi(\rho)$ will have a positive finite limit $\bar \phi(\rho)$. The formal limiting system is therefore
\begin{align}\label{eq:L}
	\begin{cases}
		\partial_t \rho + \nabla \cdot  \mathbf{q}  = \nabla \cdot ( \rho  \nabla  \bar\phi ),                                   \\
		\partial_t  \mathbf{q}  + \nabla \cdot \Big( \frac{ \mathbf{q} \otimes \mathbf{q} }{\rho} \Big) = \nabla \cdot ( \nabla  \bar\phi \otimes  \mathbf{q} ), \\
		(\rho_{\text{max}} - \rho) \bar\phi = 0,
	\end{cases}
	\quad t>0, \quad \mathbf{x} \in \Omega \subseteq {\mathbb R}^2.
\end{align}
The congestion function has now become a third incognita $\bar\phi(t,x)$ acting as a Lagrange multiplier for the constraint $\rho = \rho_{\text{max}}$. While in the region where $\rho < \rho_{\text{max}}$, System \eqref{eq:L} reduces to \eqref{eq:PGD}, in the region where $\rho = \rho_{\text{max}}$, it takes the form (setting $\rho_{\text{max}} = 1$ for simplicity): 
\begin{align}\label{eq:L_rho=1} 
	\begin{cases}
	\nabla \cdot  \mathbf{q}  = \Delta  \bar\phi, \\
		\partial_t  \mathbf{q}  + \nabla \cdot ( \mathbf{q} \otimes \mathbf{q} ) = \nabla \cdot ( \nabla  \bar\phi \otimes  \mathbf{q} ),
		\end{cases}
\end{align}
where $\Delta  \bar\phi$ denotes the Laplacian of $\bar \phi$. In particular, the second equation of \eqref{eq:L_rho=1} can be recast into 
\begin{equation}
\partial_t  \mathbf{q}  + \nabla \cdot ( \mathbf{u} \otimes \mathbf{q} ) = 0, \qquad \mathbf{u} = \mathbf{q} -  \nabla  \bar\phi. 
\label{eq:patq}
\end{equation}
In view of the first equation of \eqref{eq:L_rho=1}, we have $\nabla \cdot \mathbf{u} = 0$, so that the equation 
$$  \mathbf{q} = \mathbf{u} +  \nabla  \bar\phi, $$
appears as a Helmoltz decomposition of the vector field $\mathbf{q}$ into a divergence-free part $\mathbf{u}$ and a curl-free part $\nabla  \bar\phi$. Stated differently, $\mathbf{u}$ appears as the projection $\mathrm{Proj}_{\mathrm{div} = 0} \mathbf{q}$ of $\mathbf{q}$ on divergence-free fields. Since $\nabla \cdot \mathbf{u} = 0$, $\mathbf{u}$ can be carried out of the divergence in the first equation of \eqref{eq:patq}, so that, in the end, the equation for $\mathbf{q}$ can be written as a transport equation
$$ \partial_t  \mathbf{q}  + \big( (\mathrm{Proj}_{\mathrm{div} = 0} \mathbf{q} ) \cdot \nabla \big) \mathbf{q} = 0. $$
So, the structure of the model in the congestion region $\rho = \rho_{\text{max}}$ is a transport of $\mathbf{q}$ by its projection on divergence-free fields. 

We now highlight the analogy and difference with the two-dimensional inviscid Euler equation in vorticity formulation. This equation is written 
\begin{align*}
\begin{cases} & \partial_t \omega + (\mathbf{u} \cdot \nabla) \omega = 0, \\
& \nabla \cdot \mathbf{u} = 0, \qquad \nabla \times \mathbf{u} = \omega, 
\end{cases}
\end{align*}
where $\omega$ is a scalar (the vorticity) and $\nabla \times$ is the one-dimensional curl, which maps a vector field to a scalar field as follows: $\nabla \times \mathbf{u} = \partial_{x_1} u_2 - \partial_{x_2} u_1$ and $\mathbf{u} = (u_1,u_2)$. The $\mathbf{q}$ equation itself is 
\begin{align}\label{eq:Euler-like}
\begin{cases} & \partial_t \mathbf{q} + (\mathbf{u} \cdot \nabla) \mathbf{q} = 0, \\
& \nabla \cdot \mathbf{u} = 0, \qquad \nabla \times \mathbf{u} = \nabla \times \mathbf{q},
\end{cases} 
\end{align}
Hence, we pass from the Euler equation to the $\mathbf{q}$-equation by replacing $\omega$ by $\mathbf{q}$ in the first equation and $\omega$ by $\nabla \times \mathbf{q}$ in the second one. 
We notice that this system is more singular than the Euler one since we lose one derivative when passing from the first equation to the second one. So, the well-posedness of this system is far from being straightforward. In particular, there is no obvious form of compactness in the map sending $\mathbf{q}$ to $\mathbf{u}$.

Finally, let us point out that the conditions of the transition between uncongested and congested regions in the limit problem, as well as the motion of the interface between two such regions, are not known. In the case of the Aw-Rascle system with congestion, these conditions were investigated in \cite{BDD2008} by examining the limits as $\varepsilon \to 0$ of solutions to the Riemann problem of the original system. This approach cannot be extended to the present dissipative Aw-Rascle model because the system cannot be put in the form of a system of first order conservation laws and solutions to the Riemann problem (if they exist) are not given by simple formulas. So, the investigation of the transition between uncongested and congested regions remains an open problem at the current stage

The passage to the limit has been studied in \cite{CNP2022} in one dimension. Previously, the $\varepsilon\rightarrow 0$ limit for the rescaled Aw-Rascle model \eqref{eq:RMAR} was studied in \cite{BDD2008}. From a modelling perspective, the limit problem is not so relevant, as many of the experiments performed in Section \ref{sec:experiments} suggest that $\varepsilon$ can be thought as a scale parameter for the pedestrians; it should be small, but not zero. Nevertheless, we correctly capture the $\varepsilon\rightarrow 0$ limit numerically in Section \ref{sec:limit}, which will serve as a good test for the robustness of the numerical schemes developed in Section \ref{sec:scheme}.

\subsection{Congestion and the fundamental diagram}
\label{sec:fundamental_diagram}

The \textit{fundamental diagram of pedestrian flow} (see, for instance, \cite{CSZ2017,SSK2005,Weidmann1993}) is the name given to the statistical relation between the density of pedestrians $\rho$ present in a certain environment (say, a corridor) and the flux $\mathbf{J}$ through this environment. While it varies across environments, and its quantification is avidly debated, there is general agreement on its qualitative properties.

\begin{figure}[h!]
	\centering
	\includegraphics{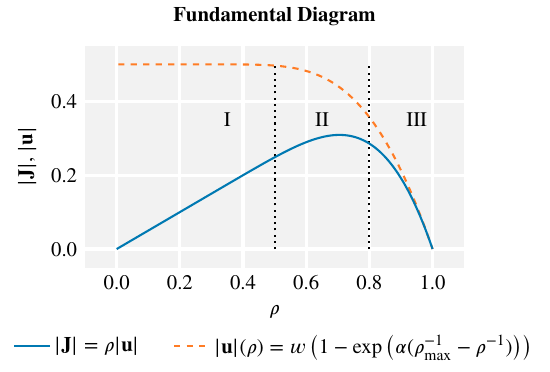}
	\caption{
		Typical form of the fundamental diagram \cite{Weidmann1993}.
		\textbf{I}: linear region.
		\textbf{II}: saturation region.
		\textbf{III}: congestion region.
		$w=0.5$, $\alpha=5$, $\rho_{\text{max}}=1$.
	}
	\label{fig:fundamental_diagram}
\end{figure}

Fig. \ref{fig:fundamental_diagram} presents a typical sketch of the fundamental diagram. The diagram is split into three regions: from lower to higher density, the \textit{linear}, \textit{saturation}, and \textit{congestion} regions. In the linear region, the magnitude of the flux grows linearly with the density; if the corridor is relatively empty, twice as many people results in twice the flux. In the saturation region, the finite capacity of the environment becomes noticeable, and the magnitude of the flux increases sub-linearly with the density, until the \textit{saturation point}, where the flux reaches its maximum value. In the congestion region, as a result of the congestion effects, the magnitude of the flux actually decreases as the density increases further.

The ability to reproduce the fundamental diagram, at least qualitatively, is the litmus test for the validity of a macroscopic pedestrian model. The congestion term in model \eqref{eq:H} will precisely address this feature: as we demonstrate in Section \ref{sec:corridor}, the model correctly captures the fundamental diagram, thereby outperforming many models which cannot.

\subsection{Derivation from an active-particle system}
\label{sec:deriv_particles}

In this section, we show that System \eqref{eq:H} can be formally derived from a system of interacting active particles. Suppose that we are given a system of ${\mathcal N}$ particles (with ${\mathcal N}$ large) described by their position $\mathbf{X}_k \in {\mathbb R}^2$ and their desired velocity $\mathbf{W}_k \in {\mathbb R}^2$ with $k \in \{1, \ldots {\mathcal N} \}$ where $(\mathbf{X}_k, \mathbf{W}_k)_{k = 1, \ldots {\mathcal N}}$ are functions ot $t$ and solutions of the following system: 
\begin{align}\label{eq:part_syst}
\begin{cases}
& \displaystyle \frac{d \mathbf{X}_k}{dt} = \mathbf{V}_k, \\
& \vspace{-0.4cm} \mbox{} \\
& \displaystyle \frac{ d \mathbf{W}_k}{dt} = 0, \\
& \vspace{-0.4cm} \mbox{} \\
& \mathbf{V}_k = \mathbf{W}_k - \nabla \big( \phi (\rho^R) \big) (\mathbf{X}_k)
\end{cases}
\end{align}
with 
\begin{equation}
\rho^R(t,\mathbf{x}) = \frac{1}{{\mathcal N}R^2} \sum_{k=1}^{{\mathcal N}} \varphi \Big( \frac{|\mathbf{x}-\mathbf{X}_k|}{R} \Big). \label{eq:rhoR}
\end{equation}
Here, $\mathbf{V}_k$ is the actual velocity of the $k$-th particle and $\rho^R(t,\mathbf{x})$ is an estimate of the particle density at~$\mathbf{x}$. It counts the proportion of particles in a small cloud around point $\mathbf{x}$ defined by the smoothing profile $\varphi$: $[0,\infty) \to [0,\infty)$ such that 
\begin{equation}
\int_{{\mathbb R}^2} \varphi(|\boldsymbol{\xi}|) \, \mathrm{d\xi} = 1, 
\label{eq:norm_aver}
\end{equation}
and the averaging radius $R$. The function $\phi$ is our previously defined congestion function, given by the last equation of \eqref{eq:H}. Here, we set $\varepsilon = 1$ for simplicity. Because of the contribution of the desired velocity $\mathbf{W}_k$ to motion, the particles are self-propelled, hence active. Like in the macroscopic model, the interaction term $- \nabla \big( \phi (\rho^R) \big) (X_k)$ describes the agents' avoidance behaviour: they turn towards the directions which lower the value of the congestion $\phi (\rho^R)$. 

In the limit ${\mathcal N} \to \infty$, we can formally describe this interacting particle system by a distribution function $f(t,\mathbf{x},\mathbf{W})$. This object is such that $f(t,\mathbf{x}, \mathbf{W}) \, \mathrm{dx} \, \mathrm{dW}$ is the number of particles such that $(\mathbf{X}_k,\mathbf{W}_k)$ belongs to a small volume of size $\mathrm{dx} \, \mathrm{dW}$ about the point $(\mathbf{x},\mathbf{W})$ at time $t$. The distribution function is subject to the kinetic equation 
\begin{align} \label{eq:kinet}
\begin{cases} 
& \partial_t f + \nabla \cdot ( \mathbf{U}_f^R \, f) = 0, \\
& \vspace{-0.4cm} \mbox{} \\
& \mathbf{U}_f^R(t,\mathbf{x},\mathbf{W}) = \mathbf{W} - \nabla \big( \phi (\rho_f^R) \big)(t,\mathbf{x}), \\
& \vspace{-0.4cm} \mbox{} \\
& \displaystyle \rho_f^R(t,\mathbf{x}) = \frac{1}{R^2} \int_{{\mathbb R}^2 \times {\mathbb R}^2} \varphi \Big( \frac{|\mathbf{x}-\mathbf{X}_k|}{R} \Big) \, f(t,\mathbf{y}, \mathbf{W}) \, \mathrm{dy} \, \mathrm{d W}. 
\end{cases}
\end{align}
Here all gradients are with respect to the variable $\mathbf{x}$. No gradient in the variable $\mathbf{W}$ appears in~\eqref{eq:kinet} because the desired particle velocities $\mathbf{W}_k$ are constant in time. The limit $R \to 0$ of this system corresponds to the case where the averaging region used in the reconstruction of the density has very small extent. In this case, owing to \eqref{eq:norm_aver}, the limit of System \eqref{eq:kinet} when $R \to 0$ gives rise to the following system: 
\begin{align} \label{eq:kinet_lim}
\begin{cases} 
& \partial_t f + \nabla \cdot ( \mathbf{U}_f \, f) = 0, \\
& \vspace{-0.4cm} \mbox{} \\
& \mathbf{U}_f(t,\mathbf{x},\mathbf{W}) = \mathbf{W} - \nabla \big( \phi (\rho_f) \big)(t,\mathbf{x}), \\
& \vspace{-0.4cm} \mbox{} \\
& \displaystyle \rho_f(t,\mathbf{x}) = \int_{{\mathbb R}^2} f(t,\mathbf{x},\mathbf{W}) \,  \mathrm{d W}, 
\end{cases}
\end{align}
which involves the local value $\rho_f$ of the density associated with $f$.

Now, we show that System \eqref{eq:kinet_lim} has special solutions in the form of a multiphase dissipative Aw-Rascle system. Before, we introduce what a measure solution to \eqref{eq:kinet_lim} is. First, let us consider a classical solution $f$ to \eqref{eq:kinet_lim}, namely a solution that is continuously differentiable (i.e. $C^1$) for $(t,\mathbf{x},\mathbf{W}) \in {\mathbb R} \times {\mathbb R}^2 \times {\mathbb R}^2$ (we ignore the questions of the initial and boundary conditions here for simplicity) and for which the density $\rho_f$ is finite everywhere. Let now $\psi$: ${\mathbb R} \times {\mathbb R}^2 \times {\mathbb R}^2 \to {\mathbb R}$ be an arbitrary $C^1$, compactly supported function (or $C^1_c$ function in short). Multiplying the first equation of \eqref{eq:kinet_lim} by $\psi$, integrating with respect to $(t,\mathbf{x},\mathbf{W})$ and using Green's formula, we get: 
\begin{equation} 
\int_{{\mathbb R} \times {\mathbb R}^2 \times {\mathbb R}^2} f \big( \partial_t \psi + \mathbf{U}_f \cdot \nabla \psi \big) \, \mathrm{dt} \, \mathrm{dx} \, \mathrm{dW} =0. 
\label{eq_kinet_week}
\end{equation}
Conversely, if $f$ belongs to $C^1$, has an everywhere-defined density $\rho_f$, is a solution to \eqref{eq_kinet_week} for all $C^1_c$ test functions $\psi$, and is a solution to the second and third equations of \eqref{eq:kinet_lim}, then, by applying Green's formula backwards, $f$ is a classical solution to~\eqref{eq:kinet_lim}. However, \eqref{eq_kinet_week} makes sense for functions $f$ that are not $C^1$. In particular, interpreting integration as a duality bracket between measures and continuous functions with compact support (i.e. $C^0_c$ functions), we can define a measure solution to \eqref{eq_kinet_week}. Let ${\mathcal M}({\mathbb R}^2 \times {\mathbb R}^2)$ be the space of measures on ${\mathbb R}^2 \times {\mathbb R}^2$ i.e., the dual of the space $C^0_c({\mathbb R}^2 \times {\mathbb R}^2)$ equipped with the weak star topology. For $\mu \in {\mathcal M}$, we define its marginal $\rho_\mu$ with respect to $\mathbf{W}$ as a measure in ${\mathcal M}({\mathbb R}^2)$ defined for any test function $\theta \in C^0({\mathbb R}^2)$ as  
\begin{equation}  
\langle \rho_\mu, \theta \rangle_{{\mathcal M}({\mathbb R}^2), C^0_c({\mathbb R}^2)} = \langle \mu, \theta \otimes \mathbf{1} \rangle_{{\mathcal M}({\mathbb R}^4), C^0_c({\mathbb R}^4)}, 
\label{eq:duality_bracket}
\end{equation}
where $\mathbf{1}$ is the constant function equal to $1$ for all $\mathbf{W} \in {\mathbb R}^2$,  $\theta \otimes \mathbf{1}$ denotes the function ${\mathbb R}^2 \times {\mathbb R}^2 \to {\mathbb R}$, $(\mathbf{x},\mathbf{W}) \mapsto \theta(\mathbf{x}) \mathbf{1}(\mathbf{W}) = \theta(\mathbf{x}) $, and $\langle \cdot, \cdot \rangle$ are duality brackets between the spaces indicated in indices. Obviously, the duality bracket on the right-hand side of \eqref{eq:duality_bracket} is not necessarily finite. If it is finite almost everywhere, we shall say that $\mu$ has finite marginals with respect to $\mathbf{W}$ almost everywhere. Now, a measure solution to \eqref{eq:kinet_lim} is a map $\mu$: ${\mathbb R} \to {\mathcal M}$ which is continuous for the weak star topology of ${\mathcal M}$, has almost everywhere finite marginals with respect to $\mathbf{W}$ for all times and is such that 
\begin{align} \label{eq_kinet_meas}
\begin{cases}
& \int_{{\mathbb R}} \big\langle \mu(t),  \partial_t \psi + \mathbf{U}_\mu \cdot \nabla \psi  \big\rangle \, \mathrm{dt} =0, \quad \forall \psi \in C^1_c({\mathbb R} \times {\mathbb R}^2 \times {\mathbb R}^2), \\
& \vspace{-0.4cm} \mbox{} \\
& \mathbf{U}_\mu (t,\mathbf{x},\mathbf{W}) = \mathbf{W} - \nabla \big( \phi (\rho_\mu) \big)(t,\mathbf{x}), \\
& \vspace{-0.4cm} \mbox{} \\
& \displaystyle \rho_\mu= \big( \mathrm{Marginal \, w.r.t.} \,\,  \mathbf{W}\big) (\mu),  
\end{cases}
\end{align}
where $\langle \cdot,\cdot  \rangle$ is the duality bracket between ${\mathcal M}$ and $C^0_c({\mathbb R}^2 \times {\mathbb R}^2)$. Multifluid solutions to this system are provided in the following:

\begin{proposition}
Let ${\mathcal P} \in {\mathbb N} \setminus \{0\}$ be an arbitrary positive integer. For each $i \in \{1, \ldots, {\mathcal P}\}$, introduce two functions $\rho_i$: ${\mathbb R} \times {\mathbb R}^2 \to [0,\infty)$, and $\mathbf{w}_i$: ${\mathbb R} \times {\mathbb R}^2 \to {\mathbb R}^2$. If $(\rho_i, \mathbf{w}_i)_{i=1, \ldots, {\mathcal P}}$ satisfies the following system: 
\begin{align} \label{eq:multiphase}
\begin{cases} 
& \partial_t \rho_i + \nabla \cdot ( \rho_i \mathbf{u}_i) = 0, \qquad i = 1, \ldots, {\mathcal P}, \\
& \vspace{-0.4cm} \mbox{} \\
& \partial_t \mathbf{w}_i + (\mathbf{u}_i \cdot \nabla) \mathbf{w}_i = 0, \qquad i = 1, \ldots, {\mathcal P}, \\
& \vspace{-0.4cm} \mbox{} \\
& \mathbf{u}_i = \mathbf{w}_i - \nabla \big( \phi (\rho) \big), \qquad i = 1, \ldots, {\mathcal P}, \\
& \vspace{-0.4cm} \mbox{} \\
& \displaystyle \rho = \sum_{i=1}^{{\mathcal P}} \rho_i, 
\end{cases}
\end{align}
then the measure
\begin{equation}
\mu(t,\mathbf{x},\mathbf{W}) = \sum_{i=1}^{{\mathcal P}} \rho_i(t,\mathbf{x}) \, \delta\big(\mathbf{W} - \mathbf{w}_i(t,\mathbf{x}) \big)
\label{eq:kinet_multiphase}
\end{equation}
is a measure solution to the kinetic equation, i.e. a solution to \eqref{eq_kinet_meas}. 
\label{prop:kinet_special_multiphase}
\end{proposition}

\begin{remark} 
To avoid complications with the initial conditions, we have defined $\rho_i$ and $\mathbf{w}_i$ on ${\mathbb R} \times {\mathbb R}^2$. However, the proposition would easily extend to the case where we include initial conditions and define $\rho_i$ and $\mathbf{w}_i$ on $[0,\infty) \times {\mathbb R}^2$. We leave the details to the reader. 
\end{remark}

\noindent
\textbf{Proof.} First, we see that $\mu$ defined by \eqref{eq:kinet_multiphase} satisfies the continuity and finiteness of marginals requirements of a measure solution of \eqref{eq_kinet_meas}. In particular, we have 
$$ \rho_\mu = \sum_{i=1}^{\mathcal P} \rho_i = \rho, $$
so that $\mathbf{U}_\mu$ is finite everywhere. Then, inserting \eqref{eq:kinet_multiphase} into the left-hand side of \eqref{eq_kinet_meas} and denoting by ${\mathcal I}_\psi$ the resulting real number, we have 
$$ {\mathcal I}_{\psi} =  \sum_{i=1}^{{\mathcal P}} \int_{{\mathbb R} \times {\mathbb R}^2} \rho_i(t,\mathbf{x}) \Big( \partial_t \psi + \mathbf{U}_\mu \cdot \nabla \psi \Big) \big( t,\mathbf{x}, \mathbf{w}_i(t,\mathbf{x}) \big) \, \mathrm{dt} \, \mathrm{dx}. $$
Thanks to the third equation of \eqref{eq:multiphase} and the second one of \eqref{eq_kinet_meas}, we deduce that
\begin{equation} 
\mathbf{u}_i (t,\mathbf{x}) = \mathbf{U}_\mu \big( t,\mathbf{x}, \mathbf{w}_i(t,\mathbf{x}) \big). 
\label{eq:u_i_vs_Umu}
\end{equation}
Besides, we have 
\begin{equation} 
\partial_t \Big[ \psi \big( t,\mathbf{x},\mathbf{w}_i(t,\mathbf{x}) \big) \Big] = \partial_t \psi \big( t,\mathbf{x}, \mathbf{w}_i(t,\mathbf{x}) \big) + \nabla_{\mathbf{W}} \psi \big(t, \mathbf{x}, \mathbf{w}_i(t,\mathbf{x}) \big) \cdot \partial_t \mathbf{w}_i (t, \mathbf{x}), 
\label{eq:partial_t_psi}
\end{equation}
where $\nabla_{\mathbf{W}}$ denotes the gradient with respect to $\mathbf{W}$. Similarly, we have
\begin{eqnarray} 
\big(\mathbf{u}_i (t,\mathbf{x}) \cdot \nabla \big) \Big[ \psi \big( t,\mathbf{x}, \mathbf{w}_i(t,\mathbf{x}) \big) \Big] &=& (\mathbf{u}_i  \cdot \nabla ) \psi \big( t,\mathbf{x},\mathbf{w}_i(t,\mathbf{x}) \big) \nonumber \\
&& 
 + \nabla_{\mathbf{W}} \psi \big( t,\mathbf{x},\mathbf{w}_i(t,\mathbf{x}) \big) \cdot \big( (\mathbf{u}_i \cdot \nabla ) \mathbf{w}_i \big) (t,\mathbf{x}) , 
\label{eq:partial_x_psi}
\end{eqnarray}
Adding \eqref{eq:partial_t_psi} and \eqref{eq:partial_x_psi}, and using \eqref{eq:u_i_vs_Umu} and the second equation of \eqref{eq:multiphase} gives 
$$ {\mathcal I}_{\psi} =  \sum_{i=1}^{{\mathcal P}} \int_{{\mathbb R} \times {\mathbb R}^2} \rho_i(t,\mathbf{x}) \, \Big( \partial_t +\mathbf{u}_i \cdot \nabla \Big) \Big[ \psi \big(t,\mathbf{x},\mathbf{w}_i(t,\mathbf{x}) \big)  \Big] \, \mathrm{dt} \, \mathrm{dx}. $$
We can now use Green's formula again and get 
$$ {\mathcal I}_{\psi} = - \sum_{i=1}^{{\mathcal P}} \int_{{\mathbb R} \times {\mathbb R}^2} \Big( \partial_t \rho_i + \nabla \cdot ( \rho_i \mathbf{u}_i) \Big) (t,\mathbf{x}) \,  \psi \big( t,\mathbf{x}, \mathbf{w}_i(t,\mathbf{x}) \big) \, \mathrm{dt} \, \mathrm{dx} = 0, $$
thanks to the first equation of~\eqref{eq:multiphase}. Hence, we have proved that $\mu$ satisfies the first equation of~\eqref{eq_kinet_meas}. Since $\mu$ satisfies the second and third equations of \eqref{eq_kinet_meas} by construction, this ends the proof. \endproof 

The dissipative Aw-Rascle system \eqref{eq:H} corresponds to a single fluid model, i.e. System~\eqref{eq:multiphase} with ${\mathcal P}=1$. This is the system we shall study in the sequel. It corresponds to a single-valued desired velocity at any point. However, multiphase models can be considered as alternatives to the single-fluid model when the latter presents issues (see Section \ref{sec:instability}). They will be investigated in future works. Finally, note that the kinetic model \eqref{eq:kinet_lim} can be viewed as a multiphase dissipative Aw-Rascle system with a continuum of desired velocities spanning the whole space~${\mathbb R}^2$. 

We finish this section with some remarks about the congestion limit $\varepsilon \to 0$ of the multifluid system \eqref{eq:multiphase} in which $\phi$ is substituted with $\varepsilon \phi$. This limit is given by the following system 
\begin{align} \label{eq:multiphase_eps=0}
\begin{cases} 
& \partial_t \rho_i + \nabla \cdot ( \rho_i \mathbf{u}_i) = 0, \qquad i = 1, \ldots, {\mathcal P}, \\
& \vspace{-0.4cm} \mbox{} \\
& \partial_t \mathbf{w}_i + (\mathbf{u}_i \cdot \nabla) \mathbf{w}_i = 0, \qquad i = 1, \ldots, {\mathcal P}, \\
& \vspace{-0.4cm} \mbox{} \\
& \mathbf{u}_i = \mathbf{w}_i - \nabla \bar \phi , \qquad i = 1, \ldots, {\mathcal P}, \\
& \vspace{-0.4cm} \mbox{} \\
& \displaystyle (\rho_{\text{max}} - \rho) \bar \phi = 0, \quad \rho = \sum_{i=1}^{{\mathcal P}} \rho_i, 
\end{cases}
\end{align}
where now, $\bar \phi = \bar \phi (t,\mathbf{x})$ is an unknown in its own right, determined by the last equation of \eqref{eq:multiphase_eps=0}. In the uncongested region where $\rho < \rho_{\text{max}}$, we have $\bar \phi = 0$ and the system reduces to a multifluid pressureless gas dynamics system
\begin{align*} 
\begin{cases} 
& \partial_t \rho_i + \nabla \cdot ( \rho_i \mathbf{w}_i) = 0, \qquad i = 1, \ldots, {\mathcal P}, \\
& \vspace{-0.4cm} \mbox{} \\
& \partial_t \mathbf{w}_i + (\mathbf{w}_i \cdot \nabla) \mathbf{w}_i = 0, \qquad i = 1, \ldots, {\mathcal P},
\end{cases}
\end{align*}
which consists of ${\mathcal P}$ uncoupled pressureless gas dynamics systems. In the congested regions where $\rho = \rho_{\text{max}}$, we can compute the equation satisfied by $\bar \phi$ by taking the sum over $i$ of the first equation of \eqref{eq:multiphase_eps=0} and using the fourth equation to remove the time derivative. After easy manipulations, we end-up with the following system, introducing $\mathbf{q}_i = \rho_i \mathbf{w}_i$: 
\begin{align} \label{eq:multiphase_congested}
\begin{cases} 
& \partial_t \rho_i + \nabla \cdot \big( \mathbf{q}_i - \rho_i\nabla \bar \phi \big) = 0, \qquad i = 1, \ldots, {\mathcal P}, \\
& \vspace{-0.4cm} \mbox{} \\
& \partial_t \mathbf{q}_i  + \nabla \cdot \Big( (\mathbf{q}_i - \rho_i \nabla \bar \phi) \otimes \frac{\displaystyle \mathbf{q}_i}{\displaystyle \rho_i} \Big) = 0, \qquad i = 1, \ldots, {\mathcal P}, \\
& \vspace{-0.4cm} \mbox{} \\
& \nabla \cdot \Big(  \sum_{i=1}^{{\mathcal P}} \mathbf{q}_i \Big) = \Delta \bar \phi. 
\end{cases}
\end{align}
This system differs significantly from the one-fluid congested System \eqref{eq:L_rho=1}. Now, $\rho_i$ cannot be eliminated and so, the system retains an equation for each of the $\rho_i$ (the first equation of \eqref{eq:multiphase_congested}). The vector $\mathbf{q}_i - \rho_i \nabla \bar \phi$ is not divergence free any longer, and so, cannot be pulled out of the divergence in the second equation of \eqref{eq:multiphase_congested}. Finally, the third equation plays the role of the first equation of \eqref{eq:L_rho=1}, but now, the terms $\nabla \cdot \mathbf{q}_i$ must be summed over $i$ to equate $\Delta \bar \phi$. So, the simple structure of System \eqref{eq:L_rho=1} which led to the Euler-like formulation \eqref{eq:Euler-like} is not preserved. One last remark is that from System \eqref{eq:multiphase_congested}, we can recover that the mass default $\zeta =: \rho_{\text{max}} - \sum_{i=1}^{\mathcal P} \rho_i$ is equal to $0$, as soon as it is equal to $0$ initially. Indeed, summing the first equation of \eqref{eq:multiphase_congested} over~$i$ and using the third equation, we have the following conservation law for $\zeta$: 
$$ \partial_t \zeta - \nabla \cdot \big( \nabla \bar \phi \, \zeta \big) = 0, $$
which has unique solution $\zeta = 0$ as soon as $\zeta = 0$ initially provided that $\nabla \bar \phi$ is smooth enough.

\setcounter{equation}{0}
\section{Numerical schemes}
\label{sec:scheme}

This section introduces structure-and-asymptotic-preserving numerical schemes for the pedestrian model proposed in Section \ref{sec:model}. We will first consider the system \eqref{eq:C} in one dimension,
\begin{align}\label{eq:C1D}
	\begin{cases}
		\partial_t \rho + \partial_x q = \varepsilon\partial_x \big( \rho\partial_x\phi (\rho) \big),                  \\
		\partial_t q +  \partial_x  \Big( \frac{q^2}{\rho} \Big) = \varepsilon \partial_x  \big(q  \partial_x \phi(\rho) \big), \\
		\phi (\rho) = \big( \rho^{-1} - \rho_{\text{max}}^{-1}\big)^{-\gamma},
	\end{cases}
	\quad t \in (0,T), \quad x \in \Omega=(0,L),
\end{align}
for $T>0$ and $L>0$, with periodic boundary conditions. The schemes will later be extended to higher dimensions through dimensional splitting.

\subsection{Stiffness issues}

The numerical simulation of \eqref{eq:C1D} is hindered by the congestion terms (those which involve $\phi$). As discussed in Section \ref{sec:model}, these terms introduce a steering behaviour in the model and enforce the capacity bound $\rho < \rho_{\text{max}}$. The bound is introduced by the singular nature of $\phi$ and its derivative, $\phi'$, which in turns render the problem stiff. The first equation may be written as
\begin{align*}
	\partial_t \rho +  \partial_x  q = \varepsilon \partial_x  \big( \rho\phi'(\rho) \partial_x \rho \big),
\end{align*}
which is a non-linear parabolic problem for the density. Observing that, in the \textit{nearly-congested} regime, $\rho_{\text{max}}-\rho \sim \varepsilon$, the term $\varepsilon \rho\phi'(\rho)$ can be expanded as
\begin{align*}
	\varepsilon \rho \phi'(\rho)
	= \varepsilon \gamma \rho^\gamma \Big( 1-\frac{\rho}{\rho_{\text{max}}}\Big)^{-(\gamma+1)}
	\sim \varepsilon^{-\gamma},
\end{align*}
we conclude that a naive numerical scheme for such a problem would require $ \Delta t \sim  \Delta x^2 / \varepsilon^{-\gamma}$ for stability, which is impractical, as $\varepsilon$ may be small.

Another potential issue is the constraint $\rho < \rho_{\text{max}}$. Not only is this a property of the solution to \eqref{eq:C1D} that we wish to capture in the numerical approximation; it is essential in order to ensure that the problem is well-defined. Unfortunately, preserving the bound numerically is non-trivial. Indeed, without the congestion terms, \eqref{eq:C} reduces to the system of pressureless gas dynamics \eqref{eq:PGD}. As discussed in Section \ref{sec:limit_explanation}, this suggests that the density has a tendency to become unbounded. While the congestion term stops Dirac deltas from appearing in \eqref{eq:C1D}, the constraint can nonetheless be hard to capture by elementary schemes. In fact, even an adaptive scheme will simply grind to a halt as $\rho$ approaches $\rho_{\text{max}}$, since the time-step required to satisfy the bound decreases faster than the solution can progress.

Our approach to circumventing these difficulties is to treat the congestion term implicitly in a manner which has previously been successful in dealing with the Euler equation with low Mach number \cite{DT2011}, as well as with a singular pressure term \cite{DHN2011}. In the remainder of this section, we introduce a first-order accurate scheme, discuss its implementation, and then generalise it to second-order spatial accuracy.

\subsection{Time semi-discrete scheme}
\label{sec:semidiscreteanalysis}

To introduce the ideas behind the schemes, we first study the semi-discrete problem, and adapt the framework from the Euler equations considered in \cite{DHN2011,DT2011}. In one dimension, the semi-discrete scheme reads
\begin{subequations}\label{eq:semidiscrete}
	\begin{align}
		\label{eq:semidiscrete_first}
		 & \frac{\rho^{n+1} - \rho^n}{ \Delta t } +  \partial_x  ( \rho^n w^n ) = \varepsilon \partial_x  ( \rho^n  \partial_x \phi^{n+1}), \\
		 & \frac{q^{n+1} - q^n}{ \Delta t } +  \partial_x  (q^n w^n) = \varepsilon \partial_x  (q^n  \partial_x \phi^{n+1}),             \\
		 & w^n = \frac{q^n}{\rho^n},                                                                     \\
		 & \phi^n = \phi(\rho^n),                                                                         \\
		 & \phi(\rho) = \big( \rho^{-1} - \rho_{\text{max}}^{-1} \big)^{-\gamma},
	\end{align}
\end{subequations}
where $x\in\Omega=(0,L)$, with periodic boundary conditions. We discretise the temporal domain uniformly, dividing it into $N+1$ equal intervals of length $ \Delta t =T/(N+1)$. The $n^{ \mathrm{th}}$ interval is $[t^n,t^{n+1})$, with $t^n=: n \Delta t $. The quantities $\rho^n$, $q^n$, $w^n$, and $\phi^n$ are, respectively, the density, the desired momentum, the desired speed, and the congestion function at time $t^n$.

This formulation immediately resolves the first numerical issue: the implicit treatment of the congestion term makes the parabolic term unconditionally stable, which will reduce the fully discrete mesh requirement to $ \Delta t \sim \Delta x$ later, a CFL condition independent of $\varepsilon$. The second issue, that of enforcing the constraint $\rho < \rho_{\text{max}}$ on the numerical solution, can also be addressed at this stage, by interpreting \eqref{eq:semidiscrete_first} as an elliptic problem for the congestion:
\begin{align}\label{eq:semidiscreteelliptic}
	-\varepsilon \Delta t  \partial_x  (\rho^n  \partial_x \phi^{n+1}) + \rho (\phi^{n+1}) = \rho^n -  \Delta t  \partial_x  ( \rho^n w^n ),
	\quad x \in \Omega=(0,L).
\end{align}
Here, $\phi \mapsto \rho(\phi)$ is the inverse function of $\rho \mapsto \phi(\rho)$; since the latter is an increasing function, so is the former, which in turn ensures the well-posedness of \eqref{eq:semidiscreteelliptic}. Provided $\rho^n\geq 0$ and $\rho^n -  \Delta t  \partial_x  (\rho^n w^n) \geq 0$ (conditions which are independent of $\varepsilon$), the maximum principle guarantees that the solution to \eqref{eq:semidiscreteelliptic} is non-negative \cite{BA1973,GT2001}. Once $\phi^{n+1}\geq 0$ is known, $\rho^{n+1}=: \rho(\phi^{n+1})$ is guaranteed to be in $[0, \rho_{\text{max}})$. Finally, because $\phi^{n+1}$ is now known, the update $q^{n+1}$ can be found explicitly.

In summary, this implicit formulation guarantees that a solution $(\rho^{n+1},q^{n+1})$ exists, and that $0 \leq \rho^{n+1} < \rho_{\text{max}}$, therefore the scheme is structure-preserving. Moreover, any requirements on $ \Delta t $ are independent of $\varepsilon$. Finally, we show below that the scheme remains a valid approximation scheme in the $\varepsilon\rightarrow 0$ limit for the limit problem \eqref{eq:L}, showing that it is asymptotic-preserving. 

Indeed, let us denote by $(\rho^n_\varepsilon, q^n_\varepsilon)_{n \in {\mathbb N}}$ the solution of \eqref{eq:semidiscrete} for a given $\varepsilon >0$ and assume that $\rho_{\textrm{max}} = 1$ for simplicity. Suppose that $(\rho^n_\varepsilon, q^n_\varepsilon) \to (\rho^n, q^n)$ when $\varepsilon \to 0$ for all $n \in {\mathbb N}$. Let $n \in {\mathbb N}$ and a position $\mathbf{x}$ be fixed. Let us examine two cases: 
\begin{itemize}
\item[-] \textbf{First case: uncongested region.} Here we assume $\rho^{n+1} < 1$. Then, $\rho^{n+1}_\varepsilon < 1$ for $\varepsilon$ small enough. Consequently $\varepsilon \phi^{n+1}_\varepsilon =: \varepsilon \phi(\rho^{n+1}_\varepsilon) \to 0$ as $\varepsilon \to 0$ so that the limit $\varepsilon \to 0$ of the scheme \eqref{eq:semidiscrete} leads to
\begin{align*}
		 & \frac{\rho^{n+1} - \rho^n}{ \Delta t } +  \partial_x  ( \rho^n w^n ) = 0, \\
		 & \frac{q^{n+1} - q^n}{ \Delta t } +  \partial_x  (q^n w^n) = 0,             
\end{align*}
with $w^n = q^n/\rho^n$. This is a time semi-discretization of \eqref{eq:L} in dimension $1$ when $\rho < 1$. 
\item[-] \textbf{Second case: congested region.} We now supppose that $\rho^n = \rho^{n+1} = 1$. Then, there exists $\bar \phi^{n+1} >0$ such that $\varepsilon \phi^{n+1}_\varepsilon \to \bar \phi^{n+1}$ as $\varepsilon \to 0$. Then, the limit $\varepsilon \to 0$ of the scheme \eqref{eq:semidiscrete} leads to
	\begin{align*}
		 & \partial_x  w^n  = \partial^2_x  \bar \phi^{n+1}, \\
		 & \frac{q^{n+1} - q^n}{ \Delta t } +  \partial_x  (q^n w^n) = \partial_x  (q^n  \partial_x \bar \phi^{n+1}),         
	\end{align*}
with now  $w^n = q^n$. This is a time semi-discretization of \eqref{eq:L} in dimension $1$ when $\rho = 1$. 
\end{itemize}
We note that the two cases above do not exhaust the list of possible cases. In particular, the case $\rho^{n+1} = 1$ and $\rho^n <1$ is not treated. This is the case where the point $\mathbf{x}$ belongs to the uncongested region at time $t^n$ and to the congested region at time $t^{n+1}$. So, $\mathbf{x}$ is traversed by the interface between the uncongested and congested regions during the time step $[t^n,t^{n+1}]$. As pointed out in Section \ref{sec:limit_explanation}, the behaviour of the transition interface between the uncongested and congested regions driven by the limit model \eqref{eq:L} is unknown. So, the analysis of this case is out-of-reach for the moment and we must skip it. However, the previous analysis shows that, at least within either the uncongested or congested regions, the time semi-discrete scheme \eqref{eq:semidiscrete} gives a valid description of the limit model \eqref{eq:L} in the limit $\varepsilon \to 0$.

\subsection{Fully discrete first-order scheme (S1)}

To treat the fully discrete problem, we discretise the spatial domain uniformly, dividing it into $M$ equal cells of length $ \Delta x=L/M$. The $i^{\mathrm{th}}$ cell is $[x_{i-1/2},x_{i+1/2})$, and its centre is $x_i$, where $x_i= \Delta x \, (i-1/2)$. The {\em discrete solution}, $(\rho_i^n, q_i^n)$, approximates the continuous solution in the finite-volume sense,
\begin{align*} 
	\rho_i^n \simeq \intbar_{x_{i-1/2}}^{x_{i+1/2}} \rho(t^n,x) \mathrm{dx},\quad
	q_i^n \simeq \intbar_{x_{i-1/2}}^{x_{i+1/2}} q(t^n,x) \mathrm{dx},
\end{align*}
where $\intbar_a^b f(s) \mathrm{ds} = \frac{1}{|b-a|} \int_a^b f(s) \mathrm{ds}$ denotes the average of $f$ over the interval $(a,b)$.

We now prescribe the implicit, first-order accurate, fully discrete scheme. The method is derived by treating the transport and the congestion terms in \eqref{eq:semidiscrete} separately; the former, with the upwind method, and the latter, through centred fluxes. The scheme reads:
\begin{subequations}\label{eq:firstorder}
\begin{align}
\label{eq:firstorderDensity}
& \frac{\rho_i^{n+1}-\rho_i^n}{ \Delta t } + \frac{F_{i+1/2}^n-F_{i-1/2}^n}{ \Delta x} = \varepsilon\frac{D_{i+1/2}^{n+1}-D_{i-1/2}^{n+1}}{ \Delta x},\\
\label{eq:firstorderMomentum}
& \frac{q_i^{n+1}-q_i^n}{ \Delta t } + \frac{G_{i+1/2}^n-G_{i-1/2}^n}{ \Delta x} = \varepsilon\frac{C_{i+1/2}^{n+1}-C_{i-1/2}^{n+1}}{ \Delta x}.
\end{align}
The system has a pair of upwind fluxes ($F$, $G$) and a pair of centred fluxes ($D$, $C$). The upwind fluxes are given by
\begin{align}
& F_{i+1/2}^n = \rho_i^n (w_{i+1/2}^n)^+ + \rho_{i+1}^n (w_{i+1/2}^n)^-,
\quad G_{i+1/2}^n = q_i^n (w_{i+1/2}^n)^+ + q_{i+1}^n (w_{i+1/2}^n)^-,
\end{align}
where $s^+=:\max \{s,0\}$ and $s^- =:\min \{s,0\}$, for any $s \in {\mathbb R}$, denote respectively the positive and negative parts of $s$. The centred fluxes are defined as
\begin{align}
& D_{i+1/2}^{n+1} = \frac{(\rho_i^n+\rho_{i+1}^n) (\phi_{i+1}^{n+1}-\phi_i^{n+1})}{2 \Delta x},
\quad C_{i+1/2}^{n+1} = \frac{(q_i^n+q_{i+1}^n) (\phi_{i+1}^{n+1}-\phi_i^{n+1})}{2 \Delta x}.
\end{align}
Furthermore, the interface velocities are defined as the average of the cell values of $w$,
\begin{align}
w_{i+1/2}^n = \frac{w_i^n+w_{i+1}^n}{2},
\end{align}
which, in turn, are defined as the ratios of momenta to densities,
\begin{align}
\quad w_i^n = \frac{q_i^n}{\rho_i^n}.
\end{align}
Finally, the congestion terms are defined by evaluating the congestion function at the cell values of the density,
\begin{align}
\quad \phi_i^n = \phi (\rho_i^n).
\end{align}
We choose periodic boundary conditions
\begin{align}
F_{1/2} = F_{M+1/2}, \quad
D_{1/2} = D_{M+1/2}, \quad
G_{1/2} = G_{M+1/2}, \quad
C_{1/2} = C_{M+1/2}.
\end{align}
\end{subequations}

As discussed in Section \ref{sec:semidiscreteanalysis}, the implicit formulation ensures that the stability requirement, $ \Delta t \sim  \Delta x$, is independent of $\varepsilon$. The constraint $\rho < \rho_{\text{max}}$ is once again enforced by not solving the density equation directly. Rather, as was done at the semi-discrete level, we define the elliptic problem
\begin{align*}
	-\varepsilon  \Delta t  \frac{D_{i+1/2}^{n+1}-D_{i-1/2}^{n+1}}{ \Delta x} +\rho (\phi_i^{n+1}) = \rho_i^n - \Delta t \frac{F_{i+1/2}^n-F_{i-1/2}^n}{ \Delta x},
\end{align*}
the discrete counterpart to \eqref{eq:semidiscreteelliptic}, where $\rho(\phi)$ is the inverse of $\phi(\rho)$. The solution, $\phi_i^{n+1}$, defines $\rho_i^{n+1}=:\rho(\phi_i^{n+1})$, and can be used in \eqref{eq:firstorderMomentum} to obtain $q_i^{n+1}$ explicitly.

As was the case for the semi-discrete equation, the fully discrete elliptic problem enjoys a comparison principle due to the monotonicity of $\rho(\phi)$. The conditions $\rho_i^n\geq 0$ and $\rho^n -  \Delta t  \Delta x^{-1} (F_{i+1/2}^n-F_{i-1/2}^n) \geq 0$ are sufficient to guarantee $\phi_i^{n+1}\geq 0$, and in turn $\rho_i^{n+1}=: \rho (\phi_i^{n+1})\in [0, \rho_{\text{max}})$. These conditions are again independent of $\varepsilon$, and indeed the scheme is asymptotic preserving.

\subsection{Fully discrete second-order scheme (S2)}

Scheme \eqref{eq:firstorder} addresses the numerical intricacies of the model \eqref{eq:C1D}. Its stability requirement, $ \Delta t \sim \Delta x$, does not depend on $\varepsilon$. Furthermore, the method of solution guarantees that the congestion constraint is satisfied.

In practice, however, the first-order scheme suffers from too much numerical diffusion, see Section \ref{sec:comparison}. While the solution to \eqref{eq:C1D} cannot ever develop shocks, the system can become close to a discontinuity as $\varepsilon$ is reduced. These ``near-shock'' profiles, while analytically smooth, are not captured well by first-order methods. In order to avoid this issue, we develop a scheme with second-order accuracy in space, which will better approximate the qualitative behaviour of the solution (see Section \ref{sec:comparison} and Figs. \ref{fig:comparison_1}, \ref{fig:comparison_2} and \ref{fig:comparison_3} in Appendix \ref{sec:supp_fig}). We remark that second-order accuracy in time could be achieved through the introduction of an intermediate step, see \cite{DHN2011,Tang2012}.

Scheme \eqref{eq:firstorder} can be reformulated to achieve second-order accuracy in space through slope-limiter techniques, see \cite{LeVeque2002}. We propose
\begin{subequations}\label{eq:secondorder}
\begin{align}
& \frac{\rho_i^{n+1}-\rho_i^n}{ \Delta t } + \frac{F_{i+1/2}^n-F_{i-1/2}^n}{ \Delta x} = \varepsilon\frac{D_{i+1/2}^{n+1}-D_{i-1/2}^{n+1}}{ \Delta x},\\
& \frac{q_i^{n+1}-q_i^n}{ \Delta t } + \frac{G_{i+1/2}^n-G_{i-1/2}^n}{ \Delta x} = \varepsilon\frac{C_{i+1/2}^{n+1}-C_{i-1/2}^{n+1}}{ \Delta x},
\end{align}
equipped this time with the second-order upwind fluxes
\begin{align}
& F_{i+1/2}^n = \rho_i^E  (w_{i+1/2}^n)^+ + \rho_{i+1}^W (w_{i+1/2}^n)^-,
\quad G_{i+1/2}^n = q_i^E (w_{i+1/2}^n)^+ + q_{i+1}^W (w_{i+1/2}^n)^-.
\end{align}
The reconstructed interface values are obtained through the linear interpolations
\begin{align}
& \rho_i^E = \rho_i^n+\frac{ \Delta x}{2}(\rho_x)_i^n, \quad \rho_i^W = \rho_i^n-\frac{ \Delta x}{2}(\rho_x)_i^n ,\\
& q_i^E = q_i^n+\frac{ \Delta x}{2}(q_x)_i^n, \quad q_i^W = q_i^n-\frac{ \Delta x}{2}(q_x)_i^n,
\end{align}
with slopes defined as
\begin{align}
& (\rho_x)_i^n = \mathrm{minmod} \Big(
	\frac{\rho_{i+1}^n-\rho_i^n}{ \Delta x},
	\frac{\rho_i^n-\rho_{i-1}^n}{ \Delta x}
\Big) ,\\
& (q_x)_i^n = \mathrm{minmod} \Big(
	\frac{q_{i+1}^n-q_i^n}{ \Delta x},
	\frac{q_i^n-q_{i-1}^n}{ \Delta x}
\Big) ,
\end{align}
where the minmod operator is given by
\begin{align}
\mathrm{minmod} (a,b) =
\begin{cases}
	\min(a,b) & \text{if } a,b > 0, \\
	\max(a,b) & \text{if } a,b < 0, \\
	0             & \text{otherwise}.
\end{cases}
\end{align}
The rest of scheme \eqref{eq:firstorder} is retained unaltered, since the remaining terms already constitute second-order accurate approximations.
\end{subequations}

Once again, the density equation must not be solved directly. We instead solve the elliptic problem for $\phi_i^{n+1}$, just as in the first-order case. The CFL condition for the second-order scheme is, as before, only due to the hyperbolic terms, and independent of $\varepsilon$.

\begin{remark}[Higher Dimensions, Other Boundary Conditions]
	In order to realistically model pedestrian dynamics, we will perform simulations of \eqref{eq:C} in two dimensions. To that end, we will extend scheme \eqref{eq:secondorder} to two dimensions via dimensional splitting. Furthermore, we will employ more interesting boundary conditions, such as no-flux, influx, and outflux conditions. These techniques are well known to those versed in numerical analysis, and have therefore been omitted for the sake of brevity. They are, however, written in full detail in Appendix \ref{sec:dimensional_splitting}.
\end{remark}

\setcounter{equation}{0}
\section{Numerical experiments}
\label{sec:experiments}

This section explores and validates model \eqref{eq:H} numerically, employing the first and second-order schemes, \eqref{eq:firstorder} and \eqref{eq:secondorder}. Interactive versions of some of the simulations presented in this section are available online \cite{ABDPrepWeb}. Videos of the simulations can be found in the permanent repository \cite{ABDPrepFig}.

\subsection{Validation of the schemes}
\label{sec:validation}

We begin the numerical experiments by validating the schemes \eqref{eq:firstorder} and \eqref{eq:secondorder}.

\subsubsection{Order of convergence}
\label{sec:convergence}

We first study the order of convergence of the schemes. While our discretisation of \eqref{eq:C} is quite natural, we must ensure that the schemes are valid even if congestion arises. To that end, we consider the problem in one dimension, over an interval $\Omega=(0,1)$ with periodic boundary conditions, and prescribe the initial data
\begin{align*}
	\begin{cases}
		\rho_0(x) = 0.7, \\
		w_0(x) = 0.5 - 0.4 \sin(2\pi x).
	\end{cases}
\end{align*}
The variability in the desired velocities will rapidly create congestion.

We solve the problem with both the first-order and second-order schemes, for $t \in (0,1)$. We choose $M=2^k$ for $4\leq k\leq 10$, thereby setting $ \Delta x=M^{-1}=2^{-k}$. Crucially, we perform the validation across a wide range of values of the asymptotic parameter, $\varepsilon \in \{1,10^{-1},10^{-2},10^{-3}\}$; this ensures that the scheme performs well across scales. The first-order scheme uses $ \Delta t = \Delta x/2$. The second-order scheme uses $ \Delta t = \Delta x^2$, in order to verify that it is indeed second-order accurate in space.

In the absence of an analytical solution, we compute relative errors. We compare the values of $\rho(1,x)$ in $L^1$ and $L^\infty$. Given the numerical solution with $M$ cells, $\rho_{M}$, and the previous numerical solution, $\rho_{M/2}$, we define the error in $L^1$ and $L^\infty$ as
\begin{align*}
	\textrm{Error}
	=
	\frac{
		\|\rho_{M/2}-\rho_{M}\|_p
	}{
		\|\rho_{M}\|_p
	}, \quad p=1, \, \infty. 
\end{align*}

\begin{figure}[h!]
	\centering
	\includegraphics{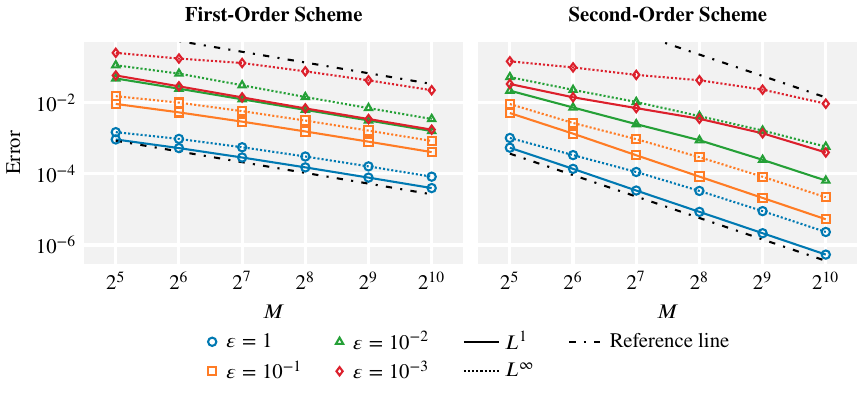}
	\caption{
		Order of convergence of the numerical schemes in the validation test of Section \ref{sec:convergence}. Error between successive numerical solutions at time $t=1$. The reference lines are $\mathcal{O}(M^{-1})$ in the first-order plot, and $\mathcal{O}(M^{-2})$ in the second-order plot.
	}
	\label{fig:validation_order}
\end{figure}

The $L^1$, and $L^\infty$ errors as a function of $M$ are shown in Fig. \ref{fig:validation_order}. Both schemes are clearly convergent, with the correct asymptotic order. A slight degeneracy can be appreciated in the second-order scheme (the $L^\infty$ norm in particular) as $\varepsilon$ approaches zero. This is, however, expected, since the density develops discontinuities in the limit, see \cite{LeVeque2002}.

\subsubsection{Comparison of first and second-order schemes}
\label{sec:comparison}

We compare the qualitative properties of the first and second-order schemes. Naturally, we expect the latter to outperform the former, given the convergence results of the previous section.

The most insightful quantity to study is the desired velocity $w$, since it satisfies an advection equation (see \eqref{eq:AD}) and thus its values should be transported along the characteristics. In particular, its extrema should be preserved exactly. However, the numerical diffusion present in the schemes smears the values of $w$. It is crucial to verify that the characteristic structure is conserved with sufficient accuracy.

We reuse the validation problem from the previous section, again setting $M=2^k$ for $4\leq k\leq 10$ and $ \Delta x=M^{-1}=2^{-k}$. We choose $\varepsilon=10^{-3}$, as it is the value that gave rise to the degeneracy in the previous section. We let $ \Delta t = \Delta x/16$ to ensure that the time-discretisation error is comparatively small.

The results are shown in Figs. \ref{fig:comparison_1}, \ref{fig:comparison_2} and \ref{fig:comparison_3} in Appendix \ref{sec:supp_fig} (see \cite{ABDPrepWeb,ABDPrepFig} for animations). As expected, the second-order scheme shows better convergence. While other fields are resolved well by the first-order scheme, the desired velocity $w$ is not; the second-order scheme much better captures the extrema. Even at the finest resolution (with $1024$ spatial cells), the first-order scheme does not quite preserve the characteristic structure, which justifies the need for at least a second-order scheme.

\subsubsection{Singular limit behaviour}
\label{sec:limit}

We study the behaviour of the second-order scheme in the $\varepsilon\rightarrow 0$ limit. As discussed in Section~\ref{sec:limit_explanation}, we expect the model to converge to \eqref{eq:L}. Here, we verify that the scheme performs well for small values of $\varepsilon$, and that the numerical solution behaves as expected in the limit.

We once more solve the validation problem, this time with $\varepsilon=10^{-k}$ for $2\leq k\leq 5$. We set $M= \Delta x^{-1}=2^{10}$, and again choose $ \Delta t = \Delta x/16$ to minimise the time-discretisation error.

The results are shown in Fig. \ref{fig:limit} in Appendix \ref{sec:supp_fig} (see \cite{ABDPrepWeb,ABDPrepFig} for animations). The second-order scheme is able to handle the small values of $\varepsilon$, as expected from the discussion in Section~\ref{sec:scheme}. Moreover, the solutions are qualitatively correct, and indeed appear to converge to a profile compatible with \eqref{eq:L}, where either $\rho < \rho_{\text{max}}$ or $\bar\phi = 0$.

\subsection{Congestion and the numerical fundamental diagram}
\label{sec:corridor}

We now study the effect of congestion on the fundamental diagram of the model, discussed in Section \ref{sec:fundamental_diagram}. To that end, we will consider a ``corridor experiment'': an elongated rectangular domain, with appropriate boundary conditions to model unidirectional flow. Pedestrians come in from the left boundary, with prescribed, constant influx density and desired speed, and can exit the domain through the right boundary, but not through the ``walls'' which run along the top and bottom boundaries. The corridor is initially empty, but gradually fills as agents enter the domain at a constant rate.

\begin{figure}[htbp]
	\centering
	\includegraphics{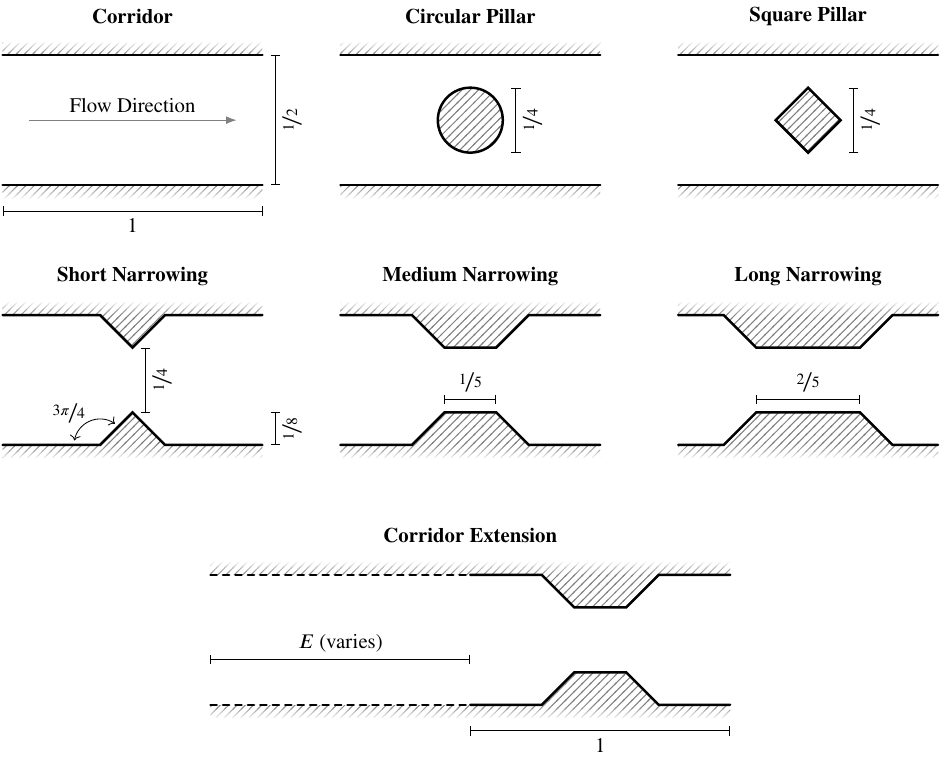}
	\caption{
		Detail of the corridor problem, obstacles, narrowings, and corridor extension.
	}
	\label{fig:corridor_diagram}
\end{figure}

As described, it is clear that the experiment reaches a steady state of uniform density and flux; however, this is no longer the case if we alter the corridor. We will explore the effects of more complex geometries, shown in Fig. \ref{fig:corridor_diagram}, which are designed to create bottlenecks, thereby allowing the study of the congestion effects.

We will first explore the effects of a ``narrowing'' in the corridor, where the walls taper inwards across some of the corridor's length. As the crowd passes through the narrowing, the effects of congestion should be visible. When the influx density is sufficiently high, we expect that further increases will result in an overall reduction of the flux through the corridor. We will also explore the effects of a ``pillar'' (an obstacle) in the middle of the corridor; a similar behaviour is expected.

We study the problem in two dimensions, over a rectangular domain $\Omega = (0,1) \times (0,0.5)$. The initial data for the empty corridor is identically zero, $\rho_0(x) = q_0(x) = 0$. We impose no-flux conditions on the top and bottom boundaries as well as outflux conditions on the right boundary. We consider influx conditions on the left boundary with a density $\rho_{\textrm{In}}$ to be specified, and a desired velocity $\mathbf{w}_{\text{In}}$ in the horizontal direction directed to the right and with magnitude $w_{\text{In}} = 0.5$, thus a desired momentum of magnitude $q_{\text{In}} = 0.5\rho_{\textrm{In}}$. For details of the implementation of influx and outflux conditions, see Appendix \ref{sec:influx}.

In practice, we find that the domain can be too small: the congestion effects that arise from the obstacles often propagate backwards far enough to reach the influx boundary, which may be conceptually inappropriate. To combat this, we consider the experiment with corridors of different lengths: we extend the domain to $\Omega = (-E,1) \times (0,0.5) $ for $E \in \{0,0.5,1,2\}$, without displacing the obstacles.

We solve the problem with the second-order scheme. Once again we let the asymptotic parameter vary across scales: $\varepsilon \in \{0.2,0.1,0.05,0.025\}$. We let $\rho_{\textrm{In}}$ increase from $0$, in increments of $0.025$, to $0.975$, in order the explore the breadth of the fundamental diagram. We choose $ \Delta x = \Delta y = 2^{-7}$, thereby setting $M_x \in \{128,192,256,384 \}$ and $M_y=64$, while we set $ \Delta t  =  \Delta x / 4$. Each simulation is run until the time $T$ where the density approximately reaches a steady state, defined here by the condition
\begin{align*}
	\frac{\| \rho(T- \Delta t )-\rho(T) \|_1}{\|\rho(T)\|_1} <10^{-6}.
\end{align*}
Note that the time $T$ is typically different for different parameter choices. See Remark \ref{rem:obstacles} of Appendix \ref{sec:influx} for details of how the obstacles are implemented in practice.

\begin{figure}[htbp]
	\centering
	\includegraphics{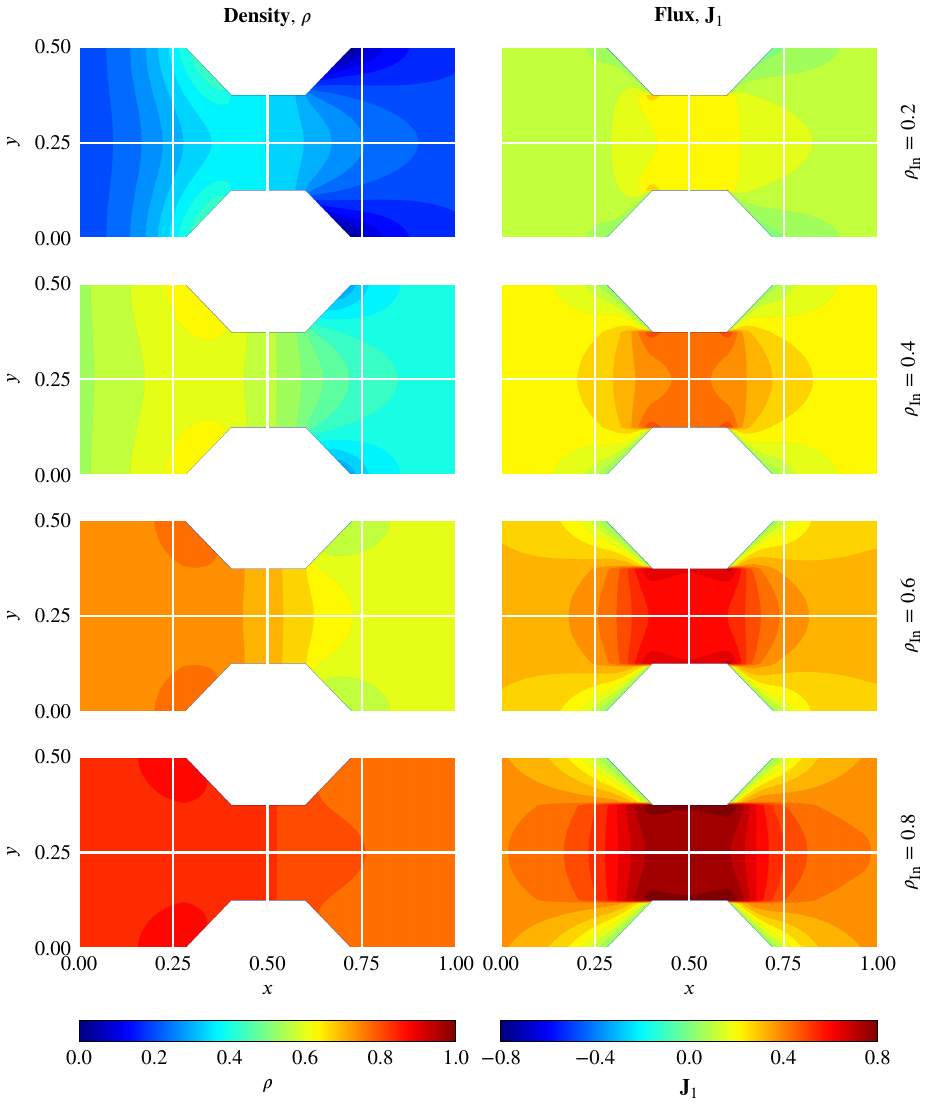}
	\caption{
		Typical steady states for the corridor narrowing test of Section \ref{sec:corridor}. \textbf{Left:} density, $\rho$. \textbf{Right:} first component of the flux, $\mathbf{J}_1$.
		$\varepsilon = 0.1$.
		$ \Delta x = \Delta y = 2^{-7}$, $M_x=384$, $M_y=64$, $ \Delta t  =  \Delta x / 4$.
		$E=2$ (corridor extension has been cropped). See \cite{ABDPrepWeb,ABDPrepFig} for animations.
	}
	\label{fig:corridor_c2_ex}
\end{figure}

\begin{figure}[htbp]
	\centering
	\includegraphics{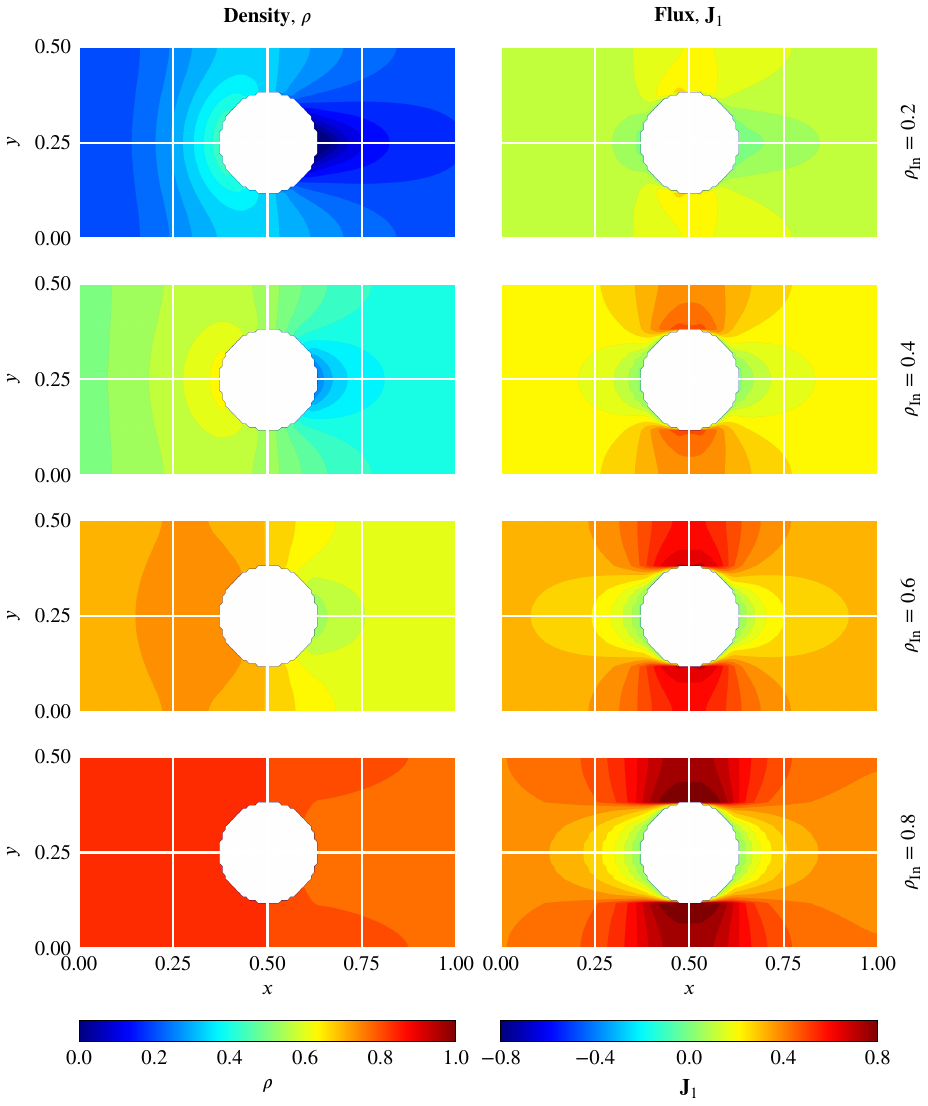}
	\caption{
		Typical steady states for the pillar obstacle test of Section \ref{sec:corridor}. \textbf{Left:} density, $\rho$. \textbf{Right:} first component of the flux, $\mathbf{J}_1$.
		$\varepsilon = 0.1$.
		$ \Delta x = \Delta y = 2^{-7}$, $M_x=384$, $M_y=64$, $ \Delta t  =  \Delta x / 4$.
		$E=2$ (corridor extension has been cropped). See \cite{ABDPrepWeb,ABDPrepFig} for animations.
	}
	\label{fig:corridor_o_ex}
\end{figure}

Figs. \ref{fig:corridor_c2_ex} and \ref{fig:corridor_o_ex} show the typical steady state reached for $\varepsilon=0.1$ and $\rho_{\textrm{In}} \in \{0.2,0.4,0.6,0.8\}$. As expected, the value of the density $\rho$ is generally higher as the crowd approaches the narrowing/obstacle, and highest just before reaching it. Upon reaching the congestion, the crowd successfully steers around it, ``spilling'' to the sides. The main difference between the two families of experiments is that the narrowing experiments concentrate the crowd in the centre of the corridor, whereas the obstacle experiments split it in half. As the obstacle is passed, the crowd progressively spreads across the full width of the corridor again. Note that the magnitude of the variation of the density in each experiment inversely correlates with $\rho_{\textrm{In}}$; when the crowd density is very high, a slight local increase constitutes significant congestion, which is not the case when the corridor is mostly empty. The horizontal component of the flux $\mathbf{J}_1$ appears to be zero just before the obstacle, and is highest just either side of the obstacle, indicating that the congested crowd is moving around the obstacle. We remark that no oscillations appear in the simulations. We will see a different situation where oscillations appear in  Section \ref{sec:instability}.

To compute the numerical fundamental diagram, we measure the average value of the first component of the flux $\mathbf{J}_1$ through the right boundary, which we denote $J_{\textrm{Eq}}(\rho_{\textrm{In}})$. The precise definition of this term is given in Remark \ref{rem:flux_corridor} of Appendix \ref{sec:dimensional_splitting}. In the absence of congestion effects, we would expect $J_{\textrm{Eq}} = q_{\textrm{In}} = 0.5 \rho_{\textrm{In}}$; however, the presence of obstacles introduces saturation and congestion effects when the background density is sufficiently high.

\begin{figure}[htbp]
	\centering
	\includegraphics{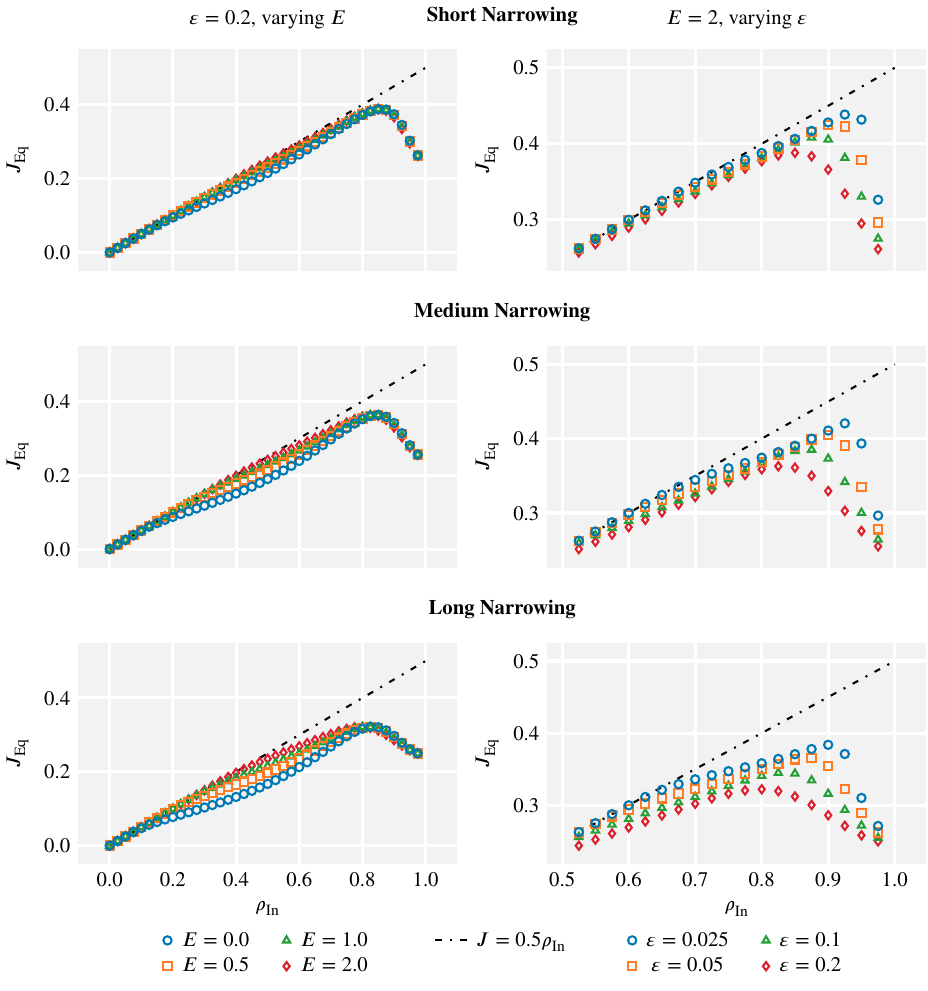}
	\caption{
		Fundamental diagram for the corridor narrowing test of Section \ref{sec:corridor}. Equilibrium flux $J_{\textrm{Eq}}$ as a function of the influx density $\rho_{\textrm{In}}$, computed at the steady state. \textbf{Left:} asymptotic parameter $\varepsilon$ is fixed, corridor extension $E$ varies. \textbf{Right:} $E$ fixed, $\varepsilon$ varies. The reference line corresponds to the free flow in a corridor without narrowings.
	}
	\label{fig:corridor_c}
\end{figure}

\begin{figure}[htbp]
	\centering
	\includegraphics{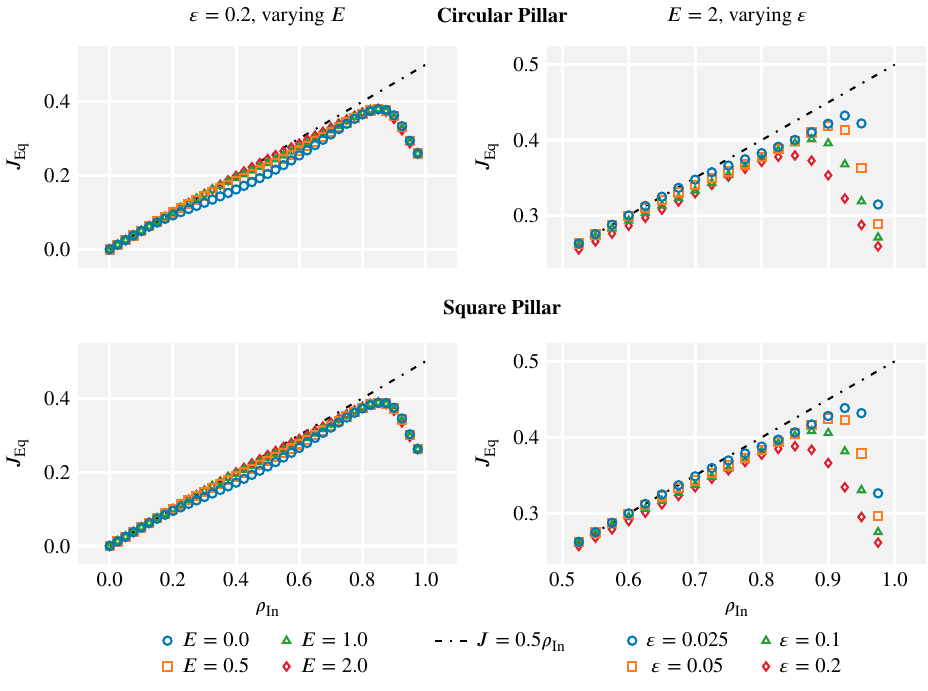}
	\caption{
		Fundamental diagram for the pillar obstacle test of Section \ref{sec:corridor}. Equilibrium flux $J_{\textrm{Eq}}$ as a function of the influx density $\rho_{\textrm{In}}$, computed at the steady state. \textbf{Left:} asymptotic parameter $\varepsilon$ is fixed, corridor extension $E$ varies. \textbf{Right:} $E$ fixed, $\varepsilon$ varies. The reference line corresponds to the free flow in a corridor without obstacles.
	}
	\label{fig:corridor_o}
\end{figure}

Figs. \ref{fig:corridor_c} and \ref{fig:corridor_o} show the fundamental diagram $J_{\textrm{Eq}}(\rho_{\textrm{In}})$ for a range of values of $E$ and $\varepsilon$, both for the narrowing and pillar experiments. The overall trend is as described in the literature and in Section \ref{sec:fundamental_diagram}: $J_{\textrm{Eq}}$ initially grows linearly with $\rho_{\textrm{In}}$, eventually saturates, and ultimately decreases. The model thus successfully captures the main features of the fundamental diagram.

In these experiments, $\varepsilon$ seems to act as a scale parameter for the pedestrians, a measure of their granular size. As $\varepsilon$ decreases, the fundamental diagram is closer to the unconstrained trend $J_{\textrm{Eq}} = q_{\textrm{In}} = 0.5 \rho_{\textrm{In}}$ for larger values of the influx density. Our interpretation of this phenomenon is that the pedestrians are comparatively smaller, and therefore can more easily flow through the space between the obstacles.

Across the experiments, the effect of varying $E$ (thus extending the corridor) is visible only for intermediate values of the density. Crucially, $J_{\textrm{Eq}}$ does not appear to be sensitive to $E$ when $\rho_{\textrm{In}}$ is high, which is evidence to the fact that the saturation arises from the obstacles, and not as a side-effect of the boundary conditions. We can confidently conclude that the congestion effects successfully incorporate the fundamental diagram in the model.

\subsection{Crowd collision and model instability}
\label{sec:instability}

We next study the ``collision'' (frontal interaction) between two crowds travelling in opposite directions. In practice, this situation results in lane formation and we want to check if the model has the ability to reproduce this behaviour. We will consider a rectangular domain $\Omega=(0,1) \times (0,0.5)$. The initial configuration consists of two groups of pedestrians, initially supported on the sets $A=: (0,0.45) \times (0.1,0.5)$ and $B=: (0.55,1) \times (0.0,0.4)$. The data are
\begin{align*}
	\rho_0(x) = \begin{cases}
		0.7 & x \in A\cup B,
		\\0 & \text{otherwise},
	\end{cases}
\end{align*}
and
\begin{align*}
	w_0(x) = \begin{cases}
		0.5 & x \in A,
		\\-0.5 & x \in B,
		\\0 & \text{otherwise}.
	\end{cases}
\end{align*}
The configuration is rotationally symmetric about the centre of the corridor. We impose no-flux conditions on the top and bottom boundaries, and periodic conditions on the left and right boundaries. For details of the implementation of the no-flux conditions, see Appendix \ref{sec:influx}. 

We solve the problem with the second-order scheme, for $t\in(0,0.5)$. We choose extreme values of the asymptotic parameter: $\varepsilon \in \{1,10^{-4}\}$. We choose $M_x = 2^k$ for $7\leq k \leq 9$, leading to $ \Delta x=M^{-1}=2^{-k}$. We set $\Delta y= \Delta x$, thereby $M_y = 2^{k-1}$, and $ \Delta t  =  \Delta x / 16$ in order to minimise the error arising from the time discretisation.

\begin{figure}[h!]
	\centering
	\includegraphics{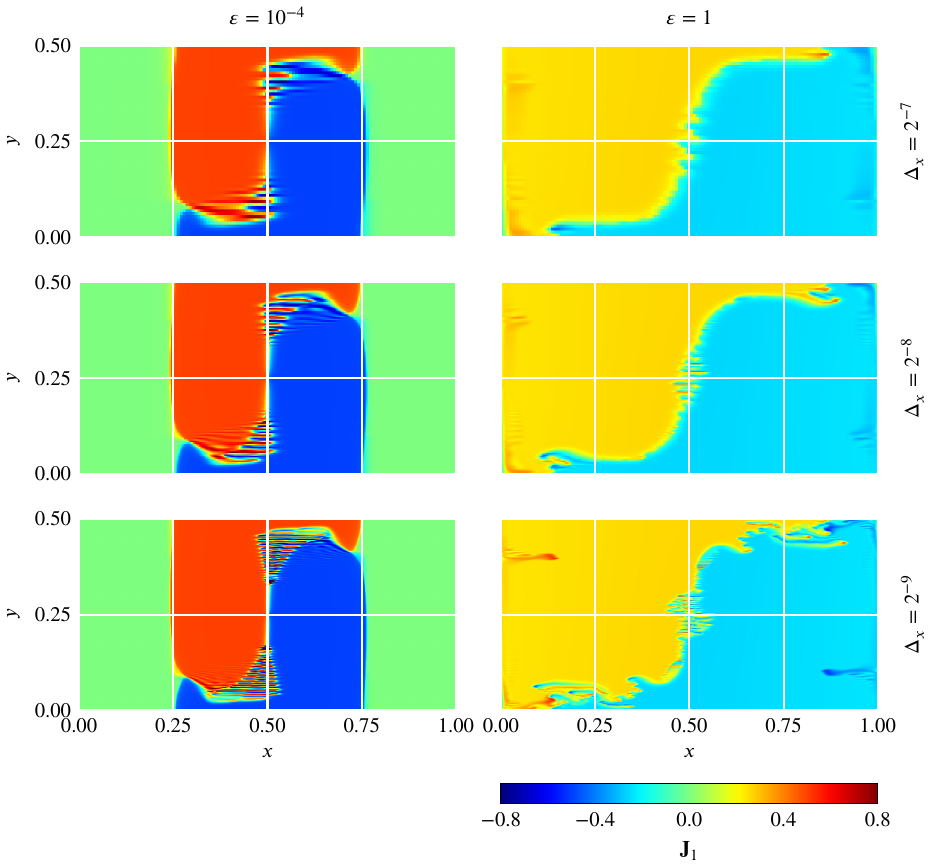}
	\caption{
	Typical instability arising in the crowd collision test of Section \ref{sec:instability}. First component of the flux, $\mathbf{J}_1$, at time $t=0.5$.
	\textbf{Left:} $\varepsilon=10^{-4}$. \textbf{Right:} $\varepsilon=1$.
	$\Delta y =  \Delta x$, $ \Delta t  =  \Delta x / 16$. See \cite{ABDPrepWeb,ABDPrepFig} for animations.
	}
	\label{fig:instability}
\end{figure}

Fig. \ref{fig:instability} shows the first component of the total flux, $\mathbf{J}_1$, at the final time, for each value of $\varepsilon$ and $ \Delta x$. Leaving aside any modelling considerations, it is clear that the numerical results are not convergent, as oscillations whose wavelengths are of the same order as the mesh size appear at the interface between the left and right-going groups. Although this is akin to lane formation, the fact that the lane width is dictated by the mesh size is clearly a flaw. In some cases, this behaviour could indicate an issue with the stability of the numerical scheme. However, we believe it is the signature of a model instability, similar to the Kelvin-Helmholtz instability of classical fluids. 
In classical fluids, as observed by Kolmogorov in his famous energy cascade,  instabilities develop increasingly smaller scales (this is the so-called inertial subrange) until diffusion kicks in and dissipates even smaller scales (in the so-called viscous dissipation subrange) \cite{chorin2013vorticity}. We could expect a similar behaviour if viscosity (i.e. diffusion) was added to the right-hand side of the $\mathbf{w}$-equation (the second equation of \eqref{eq:AD}). Unfortunately, our simulations including viscosity to the $\mathbf{w}$-equation (not shown) indicate that all structures are eventually wiped out, leading to a vanishing $\mathbf{w}$.  This does not correspond to lane-formation as observed in realistic situations. Thus, the model requires more sophisticated improvements than the mere addition of viscosity. One possible direction is to consider two-fluid models in the form of \eqref{eq:multiphase}, where one fluid corresponds to the left-going pedestrians, and the other fluid to the right-going ones. This direction will be explored in future works.

We note however, that this instability does not appear on the linear stability analysis of the model about a uniform state which is given below. This supports the idea that as long as the conditions do not depart too much from uniformity, the model can still be used. This is probably the reason why the corridor experiments shown in the  previous section do not display any problematic instability. 

To perform the stability analysis, we write the model in the following form: 
\begin{align*}
	\begin{cases}
		\partial_t \rho + \nabla \cdot  (\rho \mathbf{u}) = 0, \\
		\partial_t \mathbf{w} + (\mathbf{u} \cdot \nabla) \mathbf{w} = 0,       \\
		\mathbf{w} = \mathbf{u} + \varepsilon \nabla \phi(\rho).
	\end{cases}
\end{align*}
Given any positive real number $\rho_0$ and any vector $\mathbf{u}_0 \in {\mathbb R}^2$, then $\rho(t,\mathbf{x}) = \rho_0$ and $\mathbf{u}(t,\mathbf{x}) = \mathbf{w}(t,\mathbf{x}) = \mathbf{u}_0$ are stationary solutions of this system. Now, for a solution $(\rho,\mathbf{u},\mathbf{w})$ close enough to this stationary solution, we have $\rho = \rho_0 + \delta \rho_1$, $\mathbf{u} = \mathbf{u}_0 + \delta \mathbf{u}_1$ and $\mathbf{w} = \mathbf{u}_0 + \delta \mathbf{w}_1$, with $\delta \ll 1$. Inserting this expansion in the system and neglecting terms of order $\delta^2$ or higher, we get the following linearized system, dropping the index `$1$' for simplicity: 
\begin{equation}
	\begin{cases}
		\partial_t \rho + \rho_0 \nabla \cdot \big( \mathbf{w} - \varepsilon \phi'_0 \nabla \rho \big) + (\mathbf{u}_0 \cdot \nabla) \rho = 0, \\
		\partial_t \mathbf{w} + (\mathbf{u}_0 \cdot \nabla) \mathbf{w} = 0        
	\end{cases}
\label{eq:linearized}
\end{equation}
with $\phi'_0 = \phi'(\rho_0)$. Looking for solutions of the form  
\begin{equation} 
\left( \begin{array}{c} \rho (t,\mathbf{x})\\ \mathbf{w} (t,\mathbf{x})\end{array} \right) = \left( \begin{array}{c} \hat \rho (\lambda,\mathbf{k})\\ \hat{\mathbf{w}} (\lambda,\mathbf{k})\end{array} \right) \, e^{i \mathbf{k} \cdot \mathbf{x} - \lambda t}, 
\label{eq:Four_Lap}
\end{equation}
we find that $(\hat \rho, \hat{\mathbf{w}}) (\lambda,\mathbf{k})$ is a solution of the following homogeneous linear system 
$$
\left( \begin{array}{ccc}
- \lambda + i \mathbf{k} \cdot \mathbf{u}_0 + \varepsilon \rho_0 \phi'_0 |k|^2 & i \rho_0 k_1 &i \rho_0 k_2 \\
0 & - \lambda +  i \mathbf{k} \cdot \mathbf{u}_0  & 0  \\
0 & 0 &  - \lambda +  i \mathbf{k} \cdot \mathbf{u}_0 \end{array} \right) 
\left( \begin{array}{c} \hat{\rho} \\ \hat w_1 \\ \hat w_2 \end{array} \right) = 0 , 
$$
with $\mathbf{k}  = (k_1,k_2)$ and $\mathbf{w} = (w_1,w_2)$. The existence of a non-trivial solution to this system requires the determinant of the matrix to be $0$, i.e. $\lambda$ to be an eigenvalue of the matrix 
$$
{\mathbb J} =: \left( \begin{array}{ccc}
i \mathbf{k} \cdot \mathbf{u}_0 + \varepsilon \rho_0 \phi'_0 |k|^2 & i \rho_0 k_1 &i \rho_0 k_2 \\
0 &  i \mathbf{k} \cdot \mathbf{u}_0  & 0  \\
0 & 0 &   i \mathbf{k} \cdot \mathbf{u}_0 \end{array} \right) .
$$
The matrix ${\mathbb J}$ being triangular, its eigenvalues are the diagonal elements. Hence, there exist a simple eigenvalue $\lambda_1$ and double eigenvalue $\lambda_2$ given by
$$ \lambda_1 = i \mathbf{k} \cdot \mathbf{u}_0 + \varepsilon \rho_0 \phi'_0 |k|^2, \qquad \lambda_2 = i \mathbf{k} \cdot \mathbf{u}_0. $$
Furthemore, it is clear that the eigenspace associated with $\lambda_2$ is two-dimensional, which implies that the matrix ${\mathbb J}$ is diagonalizable. We deduce that the system is stable. Indeed, any solution of \eqref{eq:Four_Lap} will be of the form 
$$
\left( \begin{array}{c} \rho (t,\mathbf{x})\\ \mathbf{w} (t,\mathbf{x})\end{array} \right) = 
e^{i \mathbf{k} \cdot \mathbf{x}} \, e^{- i \mathbf{k} \cdot \mathbf{u}_0 t} \, \big(  {\mathbb X}_2 + e^{- \varepsilon \rho_0 \phi'_0 |k|^2 t} {\mathbb X}_1 \big), $$ 
where ${\mathbb X}_k$ are eigenvectors of ${\mathbb J}$ associated with $\lambda_k$ for $k=1, \, 2$. It is clear that this solution remains bounded in time. Since, by inversion of the Fourier transform, any solution of the linearized system \eqref{eq:linearized} can be written as a superposition of solutions of the type \eqref{eq:Four_Lap}, this system is stable.

\setcounter{equation}{0}
\section{Conclusion and outlook}
\label{sec:outlook}

This work has explored the dissipative Aw-Rascle system in the context of pedestrian dynamics. The formulation of this model, suitable for crowds, can be derived by a slight modification of the Aw-Rascle system for road traffic. The singular terms in the model strictly enforce a capacity constraint on the crowd density, and induce rich steering behaviours. We have discretised this model using asymptotic preserving numerical schemes. These schemes are able to deal with the congestion terms efficiently, and ensure that the numerical solution to the model also respects the capacity constraint. We have performed the first numerical simulations of crowds using the dissipative Aw-Rascle system in one and two dimensions. We have demonstrated the asymptotic-preserving properties of the schemes and experimentally proved its convergence. Then, using these schemes, we have explored the features of the model. We have shown that the congestion effects result in flux lowering at high density and that the resulting fundamental diagram exhibits the correct saturation behaviour. 

A drawback of the current choice of congestion function is that it renders the model ``insensitive'' to the average velocity. Because the congestion terms only involve the gradient of a function, their effect is only apparent when there are variations on the crowd density. As a consequence, any constant density which moves in a single direction with constant speed gives rise to a solution to the model without obstacles, no matter the speed. This is clearly unrealistic, because the speeds of pedestrians have an upper bound in practice. While the current model correctly predicts a reduction of the overall walking speed due to congestion, the introduction of more sophisticated terms that also penalise unrealistic physical velocities would be of interest. Introducing such effects into the model, perhaps through the dependence of the congestion function on the local velocity, is of course feasible and will be explored in future work.

The empirical calibration of the model would also be of interest. Corridor experiments similar to those simulated in Section \ref{sec:corridor} have been performed, and data is available \cite{SSK2005}. While the choices of congestion function and asymptotic parameter explored in this work yield satisfactory results, a systematic analysis based on real data would provide further insight into the qualities and validity of the model. Learning the form of the congestion function empirically, or exploring the physical correspondence between $\varepsilon$ and the size of pedestrians would be of utmost interest.

Last, but certainly not least, the instability described in Section \ref{sec:instability} should be explored further. One option would be to modify model \eqref{eq:C} to tame the instability; however, adding terms into the system without a modelling justification seems misguided. A more interesting approach would be to use the multi-fluid model \eqref{eq:multiphase} or even more, the kinetic model \eqref{eq:kinet_lim}. This will be pursued in further work.


\appendix


\vspace{1cm}
\noindent
\begin{center}
\textbf{\LARGE Appendix} 
\end{center}

\section{Two-dimensional schemes and dimensional splitting}
\label{sec:dimensional_splitting}

In this section we present the dimensional-splitting formulation of the second-order scheme~\eqref{eq:secondorder}, which was employed in the two-dimensional simulations of Section \ref{sec:experiments}. The dimensional-splitting formulation of the first order scheme \eqref{eq:firstorder} is recovered by setting the terms $(\rho_x)_{i,j}^n$, $(q_x)_{i,j}^n$, $(\rho_y)_{i,j}^n$, and $(q_y)_{i,j}^n$ to zero.

We consider \eqref{eq:C} posed over the domain $(t, x, y) \in (0,T) \times (0,L_x) \times (0,L_y)$, and discretise each of the domains uniformly, as was done in Section \ref{sec:scheme}. The temporal dimension is divided into $N+1$ equal intervals of length $ \Delta t =T/(N+1)$, each given by $[t^n,t^{n+1/2})$, with $t^n=: n \Delta t $. The spatial dimension $(0,L_x)$ is divided into $M_x$ intervals of length $ \Delta x$, given by $[x_{i-1/2},x_{i+1/2})$, where $x_i =  \Delta x (i-1/2)$. Similarly, $(0,L_y)$ is divided into $M_y$ intervals of length $\Delta y$, given by $[y_{j-1/2},y_{j+1/2})$, where $y_j = \Delta y (j-1/2)$. The spatial domain is thus split into cells, each with corners $(x_{i-1/2}, y_{j-1/2})$, $(x_{i+1/2}, y_{j-1/2})$, $(x_{i+1/2}, y_{j+1/2})$, $(x_{i-1/2}, y_{j+1/2})$, and centre $(x_i,y_j)$.

The discrete solution will approximate the continuous solution in the finite-volume sense,
\begin{align*}
	\rho_{i,j}^n \simeq \intbar_{x_{i-1/2}}^{x_{i+1/2}} \intbar_{y_{j-1/2}}^{y_{j+1/2}} \rho(t^n,x,y) \, \mathrm{dy} \, \mathrm{dx}.
\end{align*}
It will be defined through a two step method.

\begin{itemize}
\item \textbf{Step 1: evolution in the $x$-direction}

The first step defines the intermediate update $(\rho_{i,j}^{n+1/2},  \mathbf{q}_{i,j}^{n+1/2})$ as
\begin{align*}
& \frac{\rho_{i,j}^{n+1/2}-\rho_{i,j}^n}{ \Delta t } + \frac{F_{i+1/2,j}^n-F_{i-1/2,j}^n}{ \Delta x} = \varepsilon\frac{D_{i+1/2,j}^{n+1/2}-D_{i-1/2,j}^{n+1/2}}{ \Delta x},\\
& \frac{ \mathbf{q}_{i,j}^{n+1/2}- \mathbf{q}_{i,j}^n}{ \Delta t } + \frac{\mathbf{G}_{i+1/2,j}^n-\mathbf{G}_{i-1/2,j}^n}{ \Delta x} = \varepsilon\frac{\mathbf{C}_{i+1/2,j}^{n+1/2}-\mathbf{C}_{i-1/2,j}^{n+1/2}}{ \Delta x},
\end{align*}
with the upwind fluxes
\begin{align*}
& F_{i+1/2,j}^n = \rho_{i,j}^E \big( (w_1)_{i+1/2,j}^n \big)^+ + \rho_{i+1,j}^W \big( (w_1)_{i+1/2,j}^n \big)^-,\\
& \mathbf{G}_{i+1/2,j}^n =  \mathbf{q}_{i,j}^E \big( (w_1)_{i+1/2,j}^n \big)^+ +  \mathbf{q}_{i+1,j}^W \big((w_1)_{i+1/2,j}^n \big)^-.
\end{align*}
The reconstructed values are
\begin{align*}
& \rho_{i,j}^E = \rho_{i,j}^n+\frac{ \Delta x}{2}(\rho_x)_{i,j}^n, \quad \rho_{i,j}^W = \rho_{i,j}^n-\frac{ \Delta x}{2}(\rho_x)_{i,j}^n ,\\
&  \mathbf{q}_{i,j}^E =  \mathbf{q}_{i,j}^n+\frac{ \Delta x}{2} (\mathbf{q}_x)_{i,j}^n, \quad  \mathbf{q}_{i,j}^W = \mathbf{q}_{i,j}^n-\frac{ \Delta x}{2} (\mathbf{q}_x)_{i,j}^n,
\end{align*}
with slopes defined as
\begin{align*}
& (\rho_x)_{i,j}^n = \mathrm{minmod} \Big(
	\frac{\rho_{i+1,j}^n-\rho_{i,j}^n}{ \Delta x},
	\frac{\rho_{i,j}^n-\rho_{i-1,j}^n}{ \Delta x}
\Big) ,\\
& (\mathbf{q}_x)_{i,j}^n = \mathrm{minmod} \Big(
	\frac{ \mathbf{q}_{i+1,j}^n- \mathbf{q}_{i,j}^n}{ \Delta x},
	\frac{ \mathbf{q}_{i,j}^n- \mathbf{q}_{i-1,j}^n}{ \Delta x}
\Big) ,
\end{align*}
and the \textit{minmod} limiter is applied component-wise on the momentum.
The centred fluxes are
\begin{align*}
& D_{i+1/2,j}^{n+1/2} = \frac{(\rho_{i,j}^n+\rho_{i+1,j}^n) (\phi_{i+1,j}^{n+1/2}-\phi_{i,j}^{n+1/2})}{2 \Delta x}, \\
& \mathbf{C}_{i+1/2,j}^{n+1/2} = \frac{( \mathbf{q}_{i,j}^n+ \mathbf{q}_{i+1,j}^n) (\phi_{i+1,j}^{n+1/2}-\phi_{i,j}^{n+1/2})}{2 \Delta x}.
\end{align*}
The interface velocities are
\begin{align*}
(w_1)_{i+1/2,j}^n = \frac{(w_1)_{i,j}^n+(w_1)_{i+1,j}^n}{2},
\end{align*}
where $w_1$, in turn, is defined as the first component of the desired velocity vector, $ \mathbf{w} $, given by $\mathbf{w} = \mathbf{q} / \rho$. 
Finally, the congestion is defined by cell evaluation,
\begin{align*}
\quad \phi_{i,j}^{n+1/2} = \phi (\rho_{i,j}^{n+1/2}).
\end{align*}

\item \textbf{Step 2: evolution in the $y$-direction}

The second step defines the sought update $(\rho_{i,j}^{n+1},  \mathbf{q}_{i,j}^{n+1})$:
\begin{align*}
& \frac{\rho_{i,j}^{n+1}-\rho_{i,j}^{n+1/2}}{ \Delta t } + \frac{F_{i,j+1/2}^{n+1/2}-F_{i,j-1/2}^{n+1/2}}{\Delta y} = \varepsilon\frac{D_{i,j+1/2}^{n+1}-D_{i,j-1/2}^{n+1}}{\Delta y},\\
& \frac{ \mathbf{q}_{i,j}^{n+1}- \mathbf{q}_{i,j}^{n+1/2}}{ \Delta t } + \frac{\mathbf{G}_{i,j+1/2}^{n+1/2}-\mathbf{G}_{i,j-1/2}^{n+1/2}}{\Delta y} = \varepsilon\frac{\mathbf{C}_{i,j+1/2}^{n+1}-\mathbf{C}_{i,j-1/2}^{n+1}}{\Delta y},
\end{align*}
with the upwind fluxes
\begin{align*}
& F_{i,j+1/2}^{n+1/2} = \rho_{i,j}^N \big( (w_2)_{i,j+1/2}^{n+1/2} \big)^+ + \rho_{i,j+1}^S \big( (w_2)_{i,j+1/2}^{n+1/2} \big)^-,\\
& \mathbf{G}_{i,j+1/2}^{n+1/2} =  \mathbf{q}_{i,j}^N \big( (w_2)_{i,j+1/2}^{n+1/2} \big)^+ +  \mathbf{q}_{i,j+1}^S \big( (w_2)_{i,j+1/2}^{n+1/2} \big)^-.
\end{align*}
The reconstructed values are given by
\begin{align*}
& \rho_{i,j}^N = \rho_{i,j}^{n+1/2}+\frac{\Delta y}{2}(\rho_y)_{i,j}^{n+1/2}, \quad \rho_{i,j}^S = \rho_{i,j}^{n+1/2}-\frac{\Delta y}{2}(\rho_y)_{i,j}^{n+1/2} ,\\
&  \mathbf{q}_{i,j}^N =  \mathbf{q}_{i,j}^{n+1/2}+\frac{\Delta y}{2}(\mathbf{q}_y)_{i,j}^{n+1/2}, \quad  \mathbf{q}_{i,j}^S =  \mathbf{q}_{i,j}^{n+1/2}-\frac{\Delta y}{2}(\mathbf{q}_y)_{i,j}^{n+1/2},
\end{align*}
with slopes defined as
\begin{align*}
& (\rho_y)_{i,j}^{n+1/2} = \mathrm{minmod} \Big(
	\frac{\rho_{i,j+1}^n-\rho_{i,j}^n}{\Delta y},
	\frac{\rho_{i,j}^n-\rho_{i,j-1}^n}{\Delta y}
\Big) ,\\
& (\mathbf{q}_y)_{i,j}^{n+1/2} = \mathrm{minmod} \Big(
	\frac{ \mathbf{q}_{i,j+1}^n- \mathbf{q}_{i,j}^n}{\Delta y},
	\frac{ \mathbf{q}_{i,j}^n- \mathbf{q}_{i,j-1}^n}{\Delta y}
\Big) .
\end{align*}
The centred fluxes are
\begin{align*}
& D_{i,j+1/2}^{n+1} = \frac{(\rho_{i,j}^{n+1/2}+\rho_{i,j+1}^{n+1/2}) (\phi_{i,j+1}^{n+1}-\phi_{i,j}^{n+1})}{2\Delta y}, \\
& \mathbf{C}_{i,j+1/2}^{n+1} = \frac{( \mathbf{q}_{i,j}^{n+1/2}+ \mathbf{q}_{i,j+1}^{n+1/2}) (\phi_{i,j+1}^{n+1}-\phi_{i,j}^{n+1})}{2\Delta y}.
\end{align*}
The interface velocities are
\begin{align*}
(w_2)_{i,j+1/2}^{n+1/2} = \frac{(w_2)_{i,j}^{n+1/2}+(w_2)_{i,j+1}^{n+1/2}}{2},
\end{align*}
where $(w_2)$ is defined as the second component of the desired velocity $ \mathbf{w} $. 
The congestion is finally defined by cell evaluation,
\begin{align*}
\quad \phi_{i,j}^{n+1} = \phi (\rho_{i,j}^{n+1}).
\end{align*}
\end{itemize}

Just as in the one-dimensional case, the density equations are never solved directly. The updates are computed by solving for the congestion variable first; see Section \ref{sec:scheme}.

\begin{remark}[Flux through a corridor]
	The corridor experiments of Section \ref{sec:corridor} calculate the fundamental diagram of the model in several scenarios. To that end, we need to measure the flux through the corridor $J_{\textrm{Eq}}$, which we define as the average of the first component of the flux $\mathbf{J}_1$ through the right boundary:
	\begin{align*}
		J_{\textrm{Eq}} =: \intbar_{0}^{0.5} J_1(T,1,y) \, \mathrm{dy}.
	\end{align*}
	Armed with the two-dimensional scheme, we can now define the discrete analogue of this quantity precisely:
	\begin{align*}
		J_{\textrm{Eq}} =: \frac{1}{\Delta y M_y} \sum_{j=1}^{M_y} (J_1)_{M_x+1/2,\,j} \Delta y,
	\end{align*}
	where the discrete flux is defined as
	\begin{align*}
		J_{i+1/2,j} =: F_{i+1/2,j} -\varepsilon D_{i+1/2,j},
	\end{align*}
	and where the time dependency has been omitted.
\label{rem:flux_corridor}
\end{remark}

\section{Other boundary conditions}
\label{sec:influx}

\subsection{No-flux conditions}

No-flux conditions can be easily imposed by setting $F$, $D$, $G$, and $C$ to zero along a boundary. For instance, the no-flux boundary conditions along the north and south boundary in the test of Section \ref{sec:corridor} are imposed by setting
\begin{align*}
	 & F_{i,\,1-1/2} = F_{i,\,M_y-1/2} = 0,
	\\& D_{i,\,1-1/2} = D_{i,\,M_y-1/2} = 0,
	\\& \mathbf{G}_{i,\,1-1/2} = \mathbf{G}_{i,\,M_y-1/2} = 0,
	\\& \mathbf{C}_{i,\,1-1/2} = \mathbf{C}_{i,\,M_y-1/2} = 0.
\end{align*}

\begin{remark}[Corridor obstacles]
	In practice, we use no-flux boundary conditions to handle the obstacles of the corridor experiments in Section \ref{sec:corridor}. Given an obstacle, we find the mesh cells $[x_{i-1/2},x_{i+1/2}) \times [y_{j-1/2},y_{j+1/2})$ which overlap with the obstacle, and set no-flux conditions along their boundaries. The value of the scheme on those cells is then simply ignored.
\label{rem:obstacles}
\end{remark}

\subsection{Influx conditions}
The influx conditions used in Section \ref{sec:corridor} are, essentially, a Dirichlet condition, where the value of $\rho$ and $w$ is specified at the left boundary. In our scheme, these are imposed on the left boundary by choosing
\begin{align*}
	 & F_{1-1/2,\,j}^{n+1/2} = \rho_{\textrm{In}} w_{\textrm{In}},
	\\& D_{1-1/2,\,j}^{n+1/2} = \frac{(\rho_{\textrm{In}}+\rho_{1,\,j}^n) (\phi(\rho_{\textrm{In}})-\phi_{1,\,j}^{n+1/2})}{2 \Delta x},
	\\& \mathbf{G}_{1-1/2,\,j}^{n+1/2} =
	\begin{pmatrix}
		\rho_{\textrm{In}}^2 w_{\textrm{In}}
		\\ 0
	\end{pmatrix},
	\\& \mathbf{C}_{i+1/2,j}^{n+1/2} = \left(
		\begin{pmatrix}
			\rho_{\textrm{In}} w_{\textrm{In}}
			\\ 0
		\end{pmatrix}
		+ \mathbf{q}_{1,\,j}^n \right)
	\frac{\phi(\rho_{\textrm{In}})-\phi_{1,\,j}^{n+1/2}}{2 \Delta x}.
\end{align*}

\subsection{Outflux conditions}
The outflux conditions used in Section \ref{sec:corridor} emulate a profile travelling at constant speed through the boundary. These are imposed on the right boundary by choosing
\begin{align*}
	 & F_{M_x+1/2,\,j}^{n+1/2} = \rho_{M_x,\,j}^n \Big( \frac{ F_{M_x-1/2,\,j}^{n+1/2} }{ \rho_{M_x-1,\,j}^n }\Big)^+,
	\\& D_{M_x+1/2,\,j}^{n+1/2} = \rho_{M_x,\,j}^n \Big( \frac{ D_{M_x-1/2,\,j}^{n+1/2} }{ \rho_{M_x-1,\,j}^n } \Big)+,
	\\& \mathbf{G}_{M_x+1/2,\,j}^{n+1/2} =  \mathbf{q}_{M_x,\,j}^n \Big( \frac{ \mathbf{G}_{M_x-1/2,\,j}^{n+1/2} }{  \mathbf{q}_{M_x-1,\,j}^n } \Big)^+,
	\\& \mathbf{C}_{M_x+1/2,\,j}^{n+1/2} =  \mathbf{q}_{M_x,\,j}^n \Big( \frac{ \mathbf{C}_{M_x-1/2,\,j}^{n+1/2} }{  \mathbf{q}_{M_x-1,\,j}^n } \Big)^+,
\end{align*}
where the products and quotients between vectors are understood entry-wise.

\section{Supplementary figures}
\label{sec:supp_fig}

The sequel contains figures which are referenced in the main text, but not present there for brevity.

\begin{figure}
	\centering
	\includegraphics{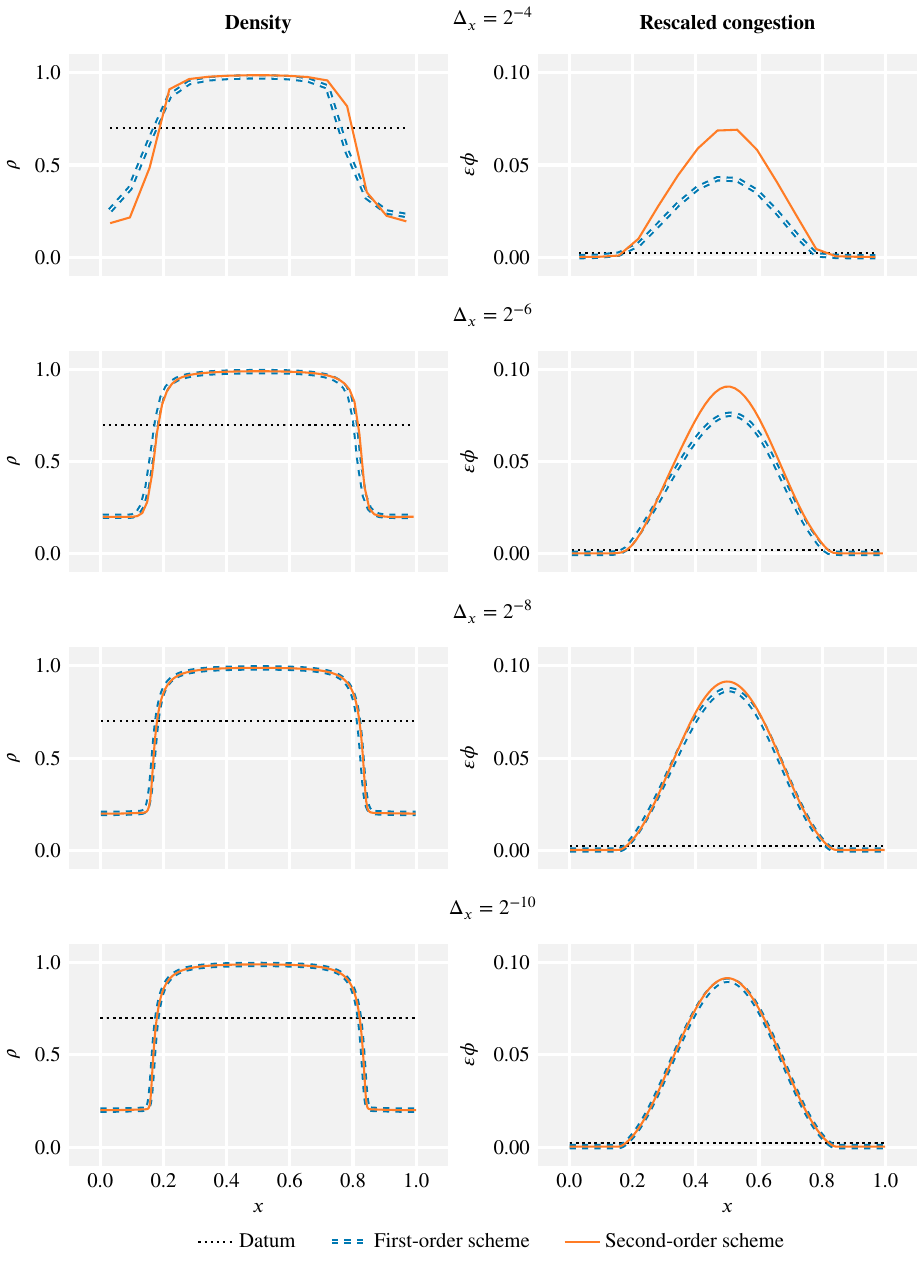}
	\caption{
	Comparison of first and second-order schemes, Section \ref{sec:comparison}.
	$\varepsilon=10^{-3}$.
	$ \Delta t = \Delta x/16$.
	See \cite{ABDPrepWeb,ABDPrepFig} for animations.
	}
	\label{fig:comparison_1}
\end{figure}

\begin{figure}
	\centering
	\includegraphics{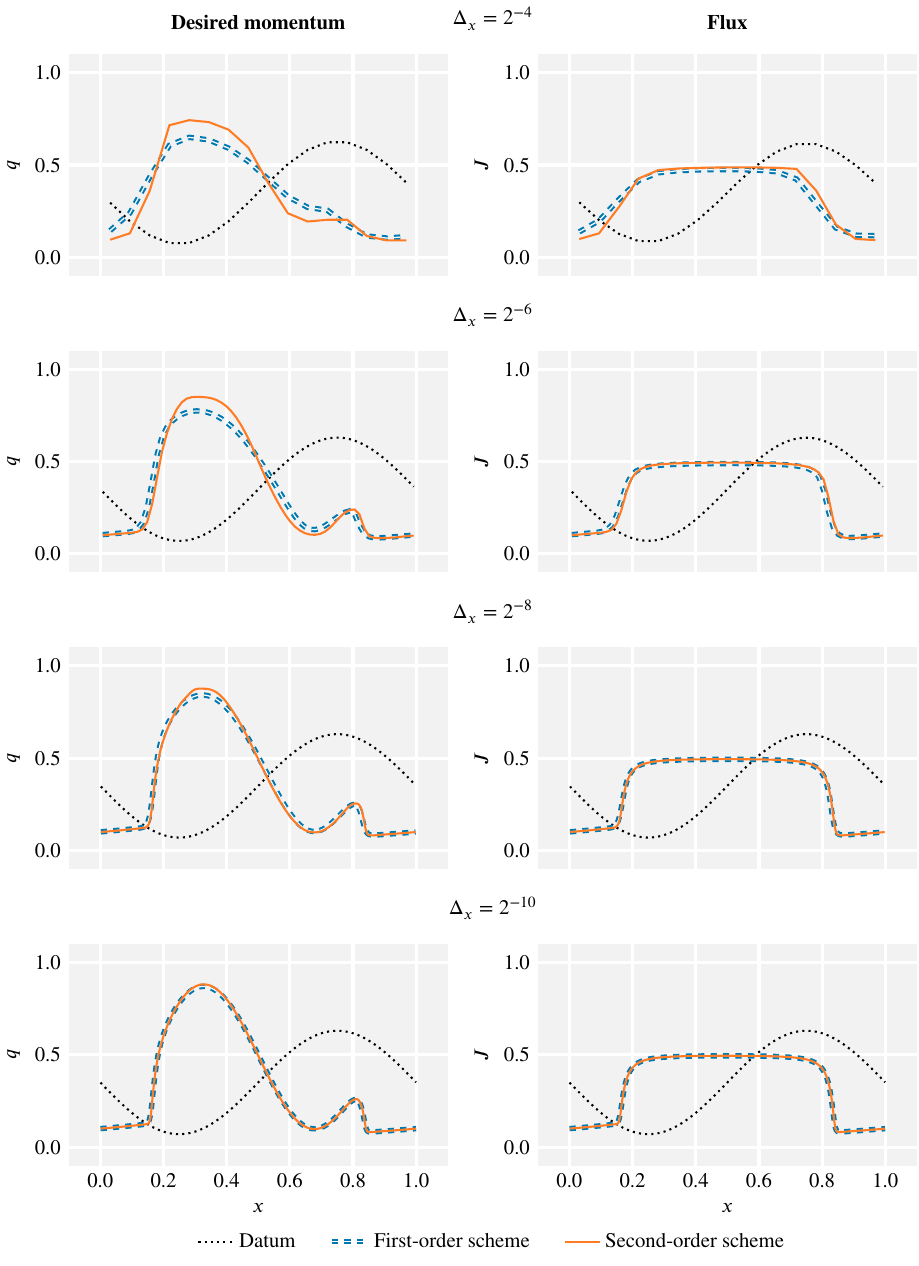}
	\caption{
	Comparison of first and second-order schemes, Section \ref{sec:comparison}.
	$\varepsilon=10^{-3}$.
	$ \Delta t = \Delta x/16$.
	See \cite{ABDPrepWeb,ABDPrepFig} for animations.
	}
\label{fig:comparison_2}
\end{figure}

\begin{figure}
	\centering
	\includegraphics{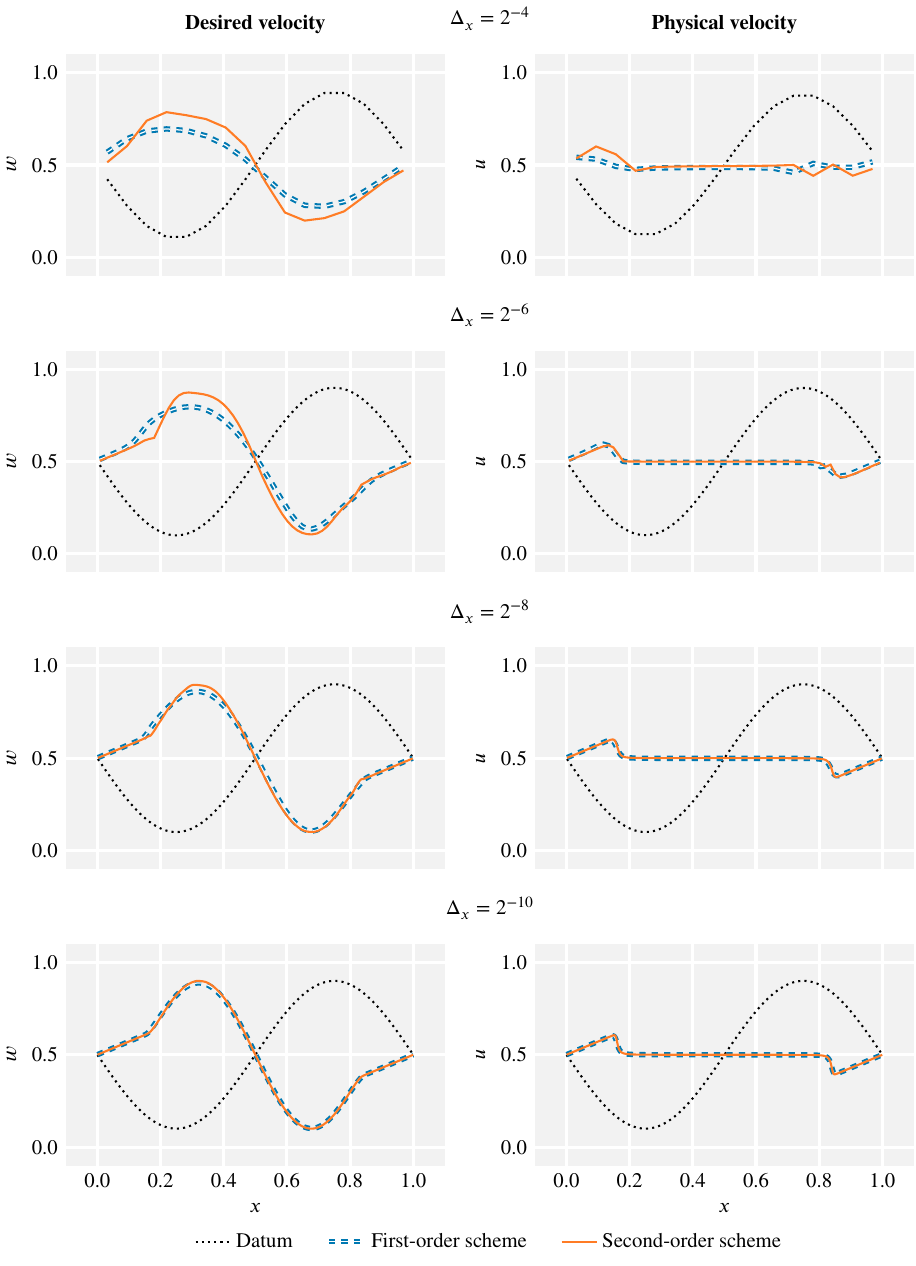}
	\caption{
	Comparison of first and second-order schemes, Section \ref{sec:comparison}.
	$\varepsilon=10^{-3}$.
	$ \Delta t = \Delta x/16$.
	See \cite{ABDPrepWeb,ABDPrepFig} for animations.
	}
\label{fig:comparison_3}
\end{figure}

\begin{figure}
	\centering
	\includegraphics{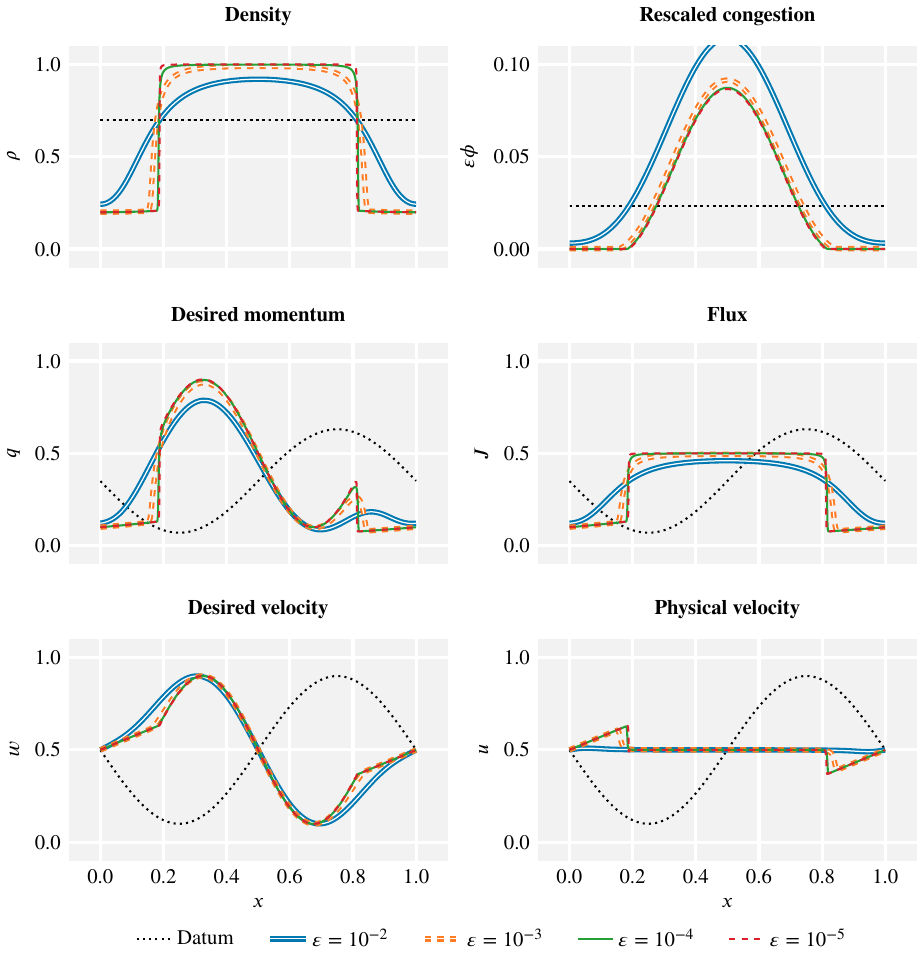}
	\caption{
	Singular limit behaviour, Section \ref{sec:limit}.
	$M= \Delta x^{-1}=2^{10}$, $ \Delta t = \Delta x/16$.
	See \cite{ABDPrepWeb,ABDPrepFig} for animations.
	}
	\label{fig:limit}
\end{figure}

\end{document}